\documentclass[twocolumn]{aastex63}

\usepackage[]{threeparttable}
\usepackage{float}
\usepackage[caption = false]{subfig}
\usepackage{natbib}
\usepackage{appendix}
\usepackage{multirow}
\usepackage{amsmath}
\usepackage[applemac]{inputenc}
\usepackage{lineno}

\usepackage{xspace}
\newcommand{\unit}[1]{\ensuremath{\mathrm{\,#1}}\xspace}
\newcommand{\kms}{\unit{km\,s^{-1}}}
\newcommand{\feh}{\unit{[Fe/H]}}
\newcommand{\code}[1]{\texttt{#1}\xspace}
\newcommand{\revise}[1]{#1}

\makeatletter
\newcommand\notsotiny{\@setfontsize\notsotiny{6}{7}}
\makeatother

\received{?}
\revised{?}
\accepted{?}
\submitjournal{The Astrophysical Journal}

\shorttitle{Stellar Spectroscopy of Three Ultra-Faint Dwarfs}
\shortauthors{Jenkins et al.}
\graphicspath{{./}{figures/}}
\begin{document}

\title{VLT Spectroscopy of Ultra-Faint Dwarf Galaxies. 1. Bo{\"o}tes I, Leo IV, Leo V }

\author[0000-0001-9827-1463]{Sydney Jenkins}
\affil{Department of Physics, University of Chicago, Chicago, IL 60637, USA}

\author[0000-0002-9110-6163]{Ting~S.~Li}
\altaffiliation{NHFP Einstein Fellow}
\affiliation{Observatories of the Carnegie Institution for Science, 813 Santa Barbara St., Pasadena, CA 91101, USA}
\affiliation{Department of Astrophysical Sciences, Princeton University, Princeton, NJ 08544, USA}
\affiliation{Department of Astronomy and Astrophysics, University of Toronto, 50 St. George Street, Toronto ON, M5S 3H4, Canada}

\author[0000-0002-6021-8760]{Andrew B. Pace}
\affiliation{McWilliams Center for Cosmology, Carnegie Mellon University, 5000 Forbes Ave, Pittsburgh, PA 15213, USA}

\author[0000-0002-4863-8842]{Alexander~P.~Ji}
\affiliation{Observatories of the Carnegie Institution for Science, 813 Santa Barbara St., Pasadena, CA 91101, USA}

\author[0000-0003-2644-135X]{Sergey~E.~Koposov}
\affiliation{Institute for Astronomy, University of Edinburgh, Royal Observatory, Blackford Hill, Edinburgh EH9 3HJ, UK}
\affiliation{Institute of Astronomy, University of Cambridge, Madingley Road, Cambridge CB3 0HA, UK}

\author[0000-0001-9649-4815]{\textsc{Bur\c{c}{\rlap{\.}\i}n Mutlu-Pakd{\rlap{\.}\i}l}}
\affiliation{Kavli Institute for Cosmological Physics, University of Chicago, Chicago, IL 60637, USA}
\affil{Department of Astronomy and Astrophysics, University of Chicago, Chicago, IL 60637, USA}

\correspondingauthor{Sydney Jenkins, Ting S. Li}
\email{sydneyjenkins@uchicago.edu, tingli@carnegiescience.edu}

\begin{abstract}
 
 We perform consistent reductions and measurements for three ultra-faint dwarf galaxies (UFDs): Bo{\"o}tes I, Leo IV and Leo V. Using the public archival data from the GIRAFFE spectrograph on the Very Large Telescope (VLT), we locate new members and provide refined measurements of physical parameters for these dwarf galaxies. We identify nine new Leo IV members and four new Leo V members, and perform a comparative analysis of previously discovered members. Additionally, we identify one new binary star in both Leo IV and Leo V. After removing binary stars, we recalculate the velocity dispersions of Bo{\"o}tes I and Leo IV to be 5.1$^{+0.7}_{-0.8}$ and 3.4$^{+1.3}_{-0.9}$ km s$^{-1}$, respectively; We do not resolve the Leo V velocity dispersion. We identify a weak velocity gradient in Leo V that is $\sim$4$\times$ smaller than the previously calculated gradient and that has a corresponding position angle which differs from the literature value by $\sim$120 deg. Combining the VLT data with previous literature, we re-analyze the Bo{\"o}tes I metallicity distribution function and find that a model including infall of pristine gas while Bo{\"o}tes I was forming stars best fits the data. Our analysis of Leo IV, Leo V and other UFDs will enhance our understanding of these enigmatic stellar populations and contribute to future dark matter studies. This is the first in a series of papers examining thirteen UDFs observed with VLT/GIRAFFE between 2009 and 2017. Similar analyses of the remaining ten UFDs will be presented in forthcoming papers.
\end{abstract}

\section{Introduction}
\label{sec:introduction}
The population of known Local Group galaxies has grown steadily in the past two decades with the discovery of many new ultra-faint dwarf (UFD) galaxies \citep[e.g.,][]{bel2006, bel2007,bel2008,bec2015,kop2015b,mau2020}. Characterized by an older stellar population, low metallicity, and low surface brightness, UFDs are the most dark-matter dominated systems known \citep{sim2007, gil2007, Simon2019ARA&A..57..375S} and may play a key role in characterizing dark matter. For instance, UFD central densities can be used to test dark matter models \citep[e.g.,][]{cal2016} and the number of Milky Way dwarf galaxy satellites can be used to constrain the mass of warm dark matter particles \citep{ken2014, Nadler2020arXiv200800022N}. Additionally, their compactness and proximity make them ideal sites for indirect dark matter detection \citep[e.g.,][]{ahn2018, Hoof2020JCAP...02..012H}.

The characteristics of individual stars within a UFD can be used to understand the global properties of the galaxy \citep[e.g.,][]{sim2007}. For example, the velocities of member stars can constrain dynamical mass and dark matter content. However, due to their low luminosity, many UFDs have few known member stars, making it difficult to provide robust measurements of the galaxies' key features. To better constrain UFDs' kinematic and metallicity parameters, we uniformly reduce and analyze archived data from the FLAMES/GIRAFFE spectrograph on the Very Large Telescopes (VLT) for thirteen UFDs. In this study, we present our membership selection process and results for three UFDs: Bo{\"o}tes I, Leo IV and Leo V. Bo{\"o}tes I data from VLT has been previously published \citep{kop2011} and is used here to validate our data reduction and membership selection processes. Our membership analysis of the remaining ten UFDs observed with VLT/GIRAFFE will be presented in a follow-up study.
 
\begin{table*}[htp!]
\centering
\tiny
\caption{VLT Ultra-Faint Dwarf Galaxies} \label{tab:table_ufd_list} 
\centering
\begin{tabular}{lllccrrrl}
\hline
\hline
 UFD & Proposal ID(s) & Time  & $N_\mathrm{exp}$ & $T_\mathrm{exp}$ (s) & NStars & Kinematic Studies \\
\hline
Bo{\"o}tes I & 182.B-0372(A) & 2/2009-3/2009 & 21 & 59350 & 118  & \citet{mun2006}, \citet{Mar2007}, \citet{lai2011}, \citet{kop2011}\\
Leo IV & 185.B-0946(A) & 5/2010-3/2011 &  17 & 45700 & 104 & \citet{sim2007}\\
Leo V & 185.B-0946(B) & 5/2010-3/2011 &  17 & 46100 & 105 & \citet{wal2009}, \citet{col2017}, \citet{mut2020} \\
\hline
Columba I & 098.B-0419(A) & 12/2016 & 6 & 16650 & 76 & \citet{fri2019} \\
Eridanus II & 096.B-0785(A) & 11/2015-12/2015 & 4 & 12000 & 110 & \citet{li2017}, \citet{zou2020}\\
Grus I & 096.B-0785(A) & 10/2015-11/2016 & 4 & 8940 & 114 & \citet{wal2016}\\
Horologium I & 096.B-0785(A) & 10/2015-10/2016 & 4 & 10980 & 114 & \citet{kop2015}\\
& 096.D-0967(B) & 12/2015-1/2016 & 14 & 38850 & 110 & \\
Horologium II & 098.B-0419(A) & 12/2016-2/2017 & 3 & 8325 & 115 & \citet{fri2019}\\
Phoenix II &096.B-0785(A) & 6/2016-9/2016 & 4 & 10980 & 105 & \citet{fri2019}\\
Reticulum II & 096.B-0785(A) & 10/2015-10/2016 & 6 & 11880 & 114 & \citet{wal2015}, \citet{sim2015}, \citet{kop2015} \\
Reticulum III & 098.B-0419(A) & 12/2016-2/2017 & 3 & 8325 & 75 & \citet{fri2019}\\ 
Segue 1 & 185.B-0946(G) & 3/2011-6/2012 & 17 & 19110 & 114 & \citet{sim2011}, \citet{geh2009} \\
& 185.B-0946(F) & 4/2011-6/2012 & 17 & 21580 & 116 & \\
Tucana II & 096.B-0785(A) & 12/2015-10/2016  & 4 & 10980 & 115 & \citet{wal2016}, \citet{chi2018}, \citet{chi2020} \\
& 096.D-0967(A) & 10/2015 & 2 & 4890 & 114 & \\
\hline
\end{tabular}
\end{table*}

Bo{\"o}tes I, Leo IV and Leo V were discovered as stellar overdensities in Sloan Digital Sky Survey (SDSS) data \citep{bel2006, bel2007, bel2008}. Bo{\"o}tes I is one of the more luminous UFDs ($M_V=-6.0$, \citealt{mun2018}). Photometric studies show that Bo{\"o}tes I is dominated by ancient metal-poor populations \citep{bro2014}, and spectroscopic studies have provided insight into the stellar kinematics and chemical abundances of Bo{\"o}tes I \citep{mun2006,Mar2007,wol2010,nor2010,kop2011,lai2011}. Additionally, \citet{kop2011} presented evidence for a two-population kinematic model, with the higher dispersion component concentrated closer to the center of the dwarf galaxy. The authors suggest that this may reflect the UFD's formation, which potentially involved the merging of several smaller populations.

\citet{sim2007} used medium-resolution spectroscopy to identify 18 members stars in Leo IV and found a velocity dispersion of 3.3$\pm$1.7 km s$^{-1}$. A follow-up analysis by \citet{kir2013} found that Leo IV has a metallicity of $-$2.29$^{+0.19}_{-0.22}$ dex and an internal metallicity spread of 0.56$^{+0.19}_{-0.14}$ dex. \citet{mun2018} have refined calculations of Leo IV's size (r$_h$ = 114$\pm$13 pc) and magnitude (M$_V$ = $-$4.99$^{+0.26}_{-0.26}$). 

Leo V is close to Leo IV in both location and radial velocity, with separations of 0.3 degrees and 50 km s$^{-1}$, respectively. These similarities have motivated investigations into their possible relationship \citep[e.g.,][]{dej2010, bla2012}). Additionally, several spectroscopic studies have targeted Leo V. \citet{wal2009} identified seven likely members, two of which are more than ten half-light radii away from Leo V's center. \citet{col2017} found an additional five members, and also presented tentative evidence of a velocity gradient (-4.1$^{+2.8}_{-2.6}$ km s$^{-1}$ per arcmin). They argue that this gradient, angled toward the Milky Way center, might suggest that Leo V is on the verge of dissolution following a close encounter with the Milky Way. Using their membership catalogue, they calculated a velocity dispersion of 2.3$^{+3.2}_{-1.6}$ km s$^{-1}$. \citet{mut2020} assessed both spectroscopic studies using high-precision photometry, spectra, and {\it Gaia} measurements. They concluded that the debris stream identified by \citet{san2012} is composed of background galaxies and foreground stars and that the velocity gradient found by \citet{col2017} may be due to small number statistics.

We locate member stars in Bo{\"o}tes I, Leo IV and Leo V using public spectroscopic data from the Very Large Telescope (VLT). The archive data were collected with the GIRAFFE spectrograph and FLAMES fiber positioner. We describe the data and data reduction in \S\ref{sec:observations} and perform velocity and metallicity measurements in \S\ref{sec:measurements}. The analyses in both \S\ref{sec:observations} and \S\ref{sec:measurements} are applied to all thirteen UFDs listed in Table~\ref{tab:table_ufd_list}, which were observed by GIRAFFE/FLAMES between 2009 and 2017. In \S\ref{sec:members}, we identify members stars in Bo{\"o}tes I, Leo IV and Leo V. We then discuss our membership results and present our updated physical parameter values in \S\ref{sec:discussion} before concluding in \S\ref{sec:conclusion}.

\section{Observations and Data Reduction}
\label{sec:observations}
\subsection{VLT Data}
\label{sec:data}
We use the publicly available data from the ESO science archive facility\footnote{\url{https://archive.eso.org/eso/eso_archive_main.html}}. Table~\ref{tab:table_ufd_list} lists the thirteen UFDs that were observed by FLAMES/GIRAFFE \citep{pas2000} between 2009 and 2017 using the LR8 grating and covering the wavelength range from 8206 - 9400~\AA~at a spectral resolution of $R\sim6,000$. 
Most of these data were unpublished at the start of this project. Notably, Bo{\"o}tes I spectroscopy was presented by \citet{kop2011}, and we use it here to validate our results.\footnote{\citet{fri2019} also used the archived VLT data to study several UFDs. A comparison with their measurements will be discussed in a future paper.} For most UFDs, one field was observed repeatedly to obtain the desired signal-to-noise ratio (S/N) and detect binaries. 

We uniformly reduce the data and calculate radial velocity and metallicity measurements for all UFDs listed in Table~\ref{tab:table_ufd_list}. In this paper, we only perform membership selection for Bo{\"o}tes I, Leo IV and Leo V. We use 21 exposures of Bo{\"o}tes I, with 20 exposure times between 45 and 58 minutes and one exposure time of 27 minutes. Bo{\"o}tes I observations took place between February and March 2009. We use 17 exposures of Leo IV of approximately 45 minutes each. These observations took place in two groups approximately eight months apart (May to July 2010 and February to March 2011). We similarly use 17 exposures of Leo V, all of approximately 45 minutes. These observations took place in two groups approximately seven months apart (July 2011 and January to March 2012). 
\subsection{Data Reduction}
\label{sec:reduction}
\begin{figure*}
\centering
\includegraphics[width=2\columnwidth]{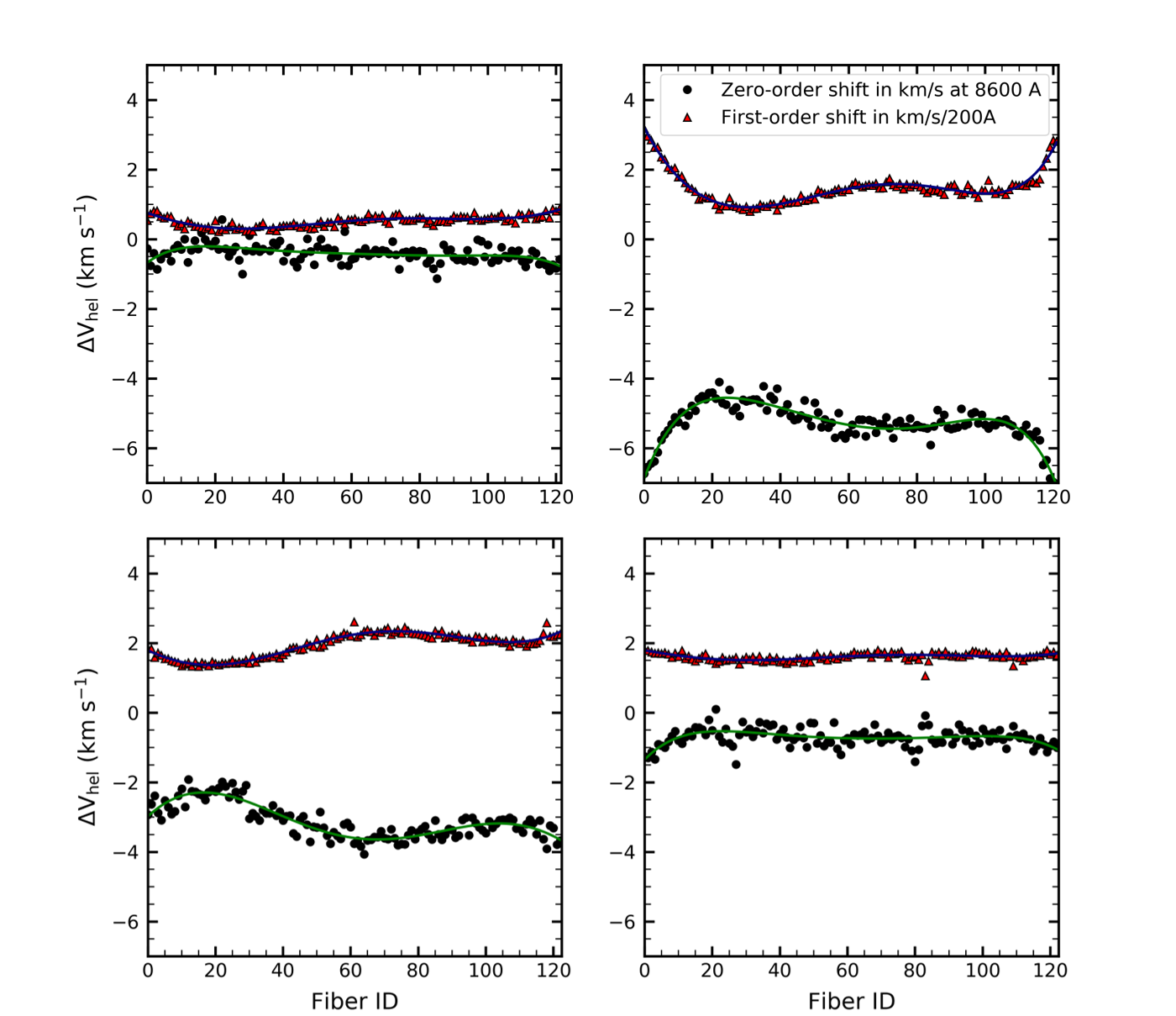}
\caption{Examples of the shifts in wavelength recalibration using sky emission lines as a function of fiber ID for each GIRAFFE exposure. The green and blue curves represent sixth degree polynomial fits to the zero and first-order shifts, respectively. We apply the shift corresponding to the polynomial fit to each spectra. Top: Two observations of Leo IV. Bottom: Two observations of Leo V. \label{specshift}}
\end{figure*}
\begin{figure}
\centering
\includegraphics[width=\columnwidth]{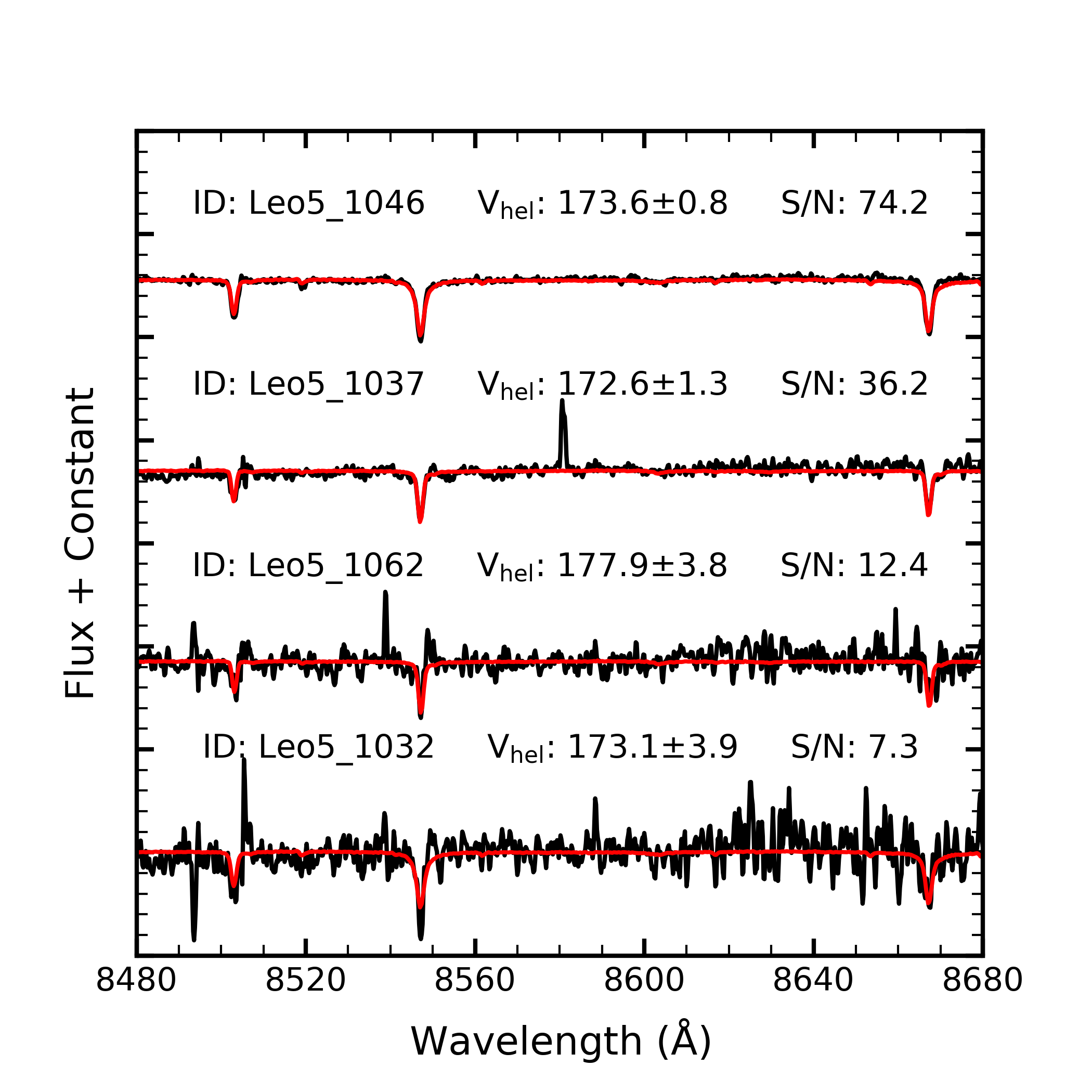}
\caption{Reduced 1D spectra of four Leo V member stars with varying S/N at wavelength range 8480--8680~\AA. The best-fit stellar templates, used to measure radial velocity, are overlaid in red. The resulting velocity measurement (in km s$^{-1}$) is provided for each spectrum, in addition to its ID and S/N. \label{examplespectra}}
\end{figure}
We reduce the raw science data associated with the studies listed in Table~\ref{tab:table_ufd_list} using the GIRAFFE Gasgano pipeline\footnote{\url{http://www.eso.org/sci/software/gasgano.html}}, which provides bias subtraction, flat-fielding, wavelength calibration and 1D spectral extraction. To minimize the fiber-dependent radial velocity offsets observed in \citet{kop2011}, the resulting spectra from each exposure are re-calibrated using sky emission lines from \citet{han2003}. \revise{A total of 12 relatively strong and isolated sky lines between 8340 \AA~and 8950 \AA~ are used in this re-calibration. We fit a linear relation between the measured wavelength $\lambda'$ and the literature value of the sky emission lines $\lambda$:}
\revise{
\begin{equation}
\frac{\lambda'-\lambda}{\lambda}c = a_1 \times \frac{\lambda-8600\mathrm{\AA}}{200\mathrm{\AA}} + a_0
\end{equation}
}

\noindent\revise{where c is the speed of light, $a_0$ is the shift in \kms at 8600 \AA~(zero-order shift), and $a_1$ is the shift for every 200 \AA~from 8600 \AA~ (the first-order shift).}
We then fit a six-degree polynomial to the zero and first-order shifts as a function of fiber number and apply the shift from the polynomial fit to each spectra. Examples of the shifts are given in Figure~\ref{specshift}, where we show the zero and first order shifts for two exposures of Leo IV and Leo V. Trends in the zero and first order shifts vary between exposures and UFDs. 

We perform sky subtraction by selecting the sky fibers in each observation and averaging them to produce a master sky spectrum. The master sky spectrum is then scaled to match the amplitude of the bright, isolated sky lines of each fiber before being subtracted from the science spectrum. We combine the individual exposures using inverse-variance weighting, creating a combined spectrum for each star. Example spectra ranging from low to high S/N are provided in Figure~\ref{examplespectra}.

To obtain parallax and proper motion data, we cross-match the VLT stars with {\it Gaia} EDR3 \citep{pru2016, bro2020} by identifying the nearest neighbor for each VLT source with separation $<$ 1". We similarly cross-match the VLT stars with the Dark Energy Camera Legacy Survey (DECaLS) DR8 \citep{Dey:2018} to obtain photometric data. All photometric data reported in this paper are reddening corrected using extinction map from \citet{Schlegel:1998}.  The proper motion and photometric data are then combined with the VLT data to form a joint catalogue of stars. In this paper, we use the IDs in the raw science data to distinguish stars, with the exception of stars in Bo{\"o}tes I; for Bo{\"o}tes I, we label stars by appending the ID from \citet{kop2011} to the prefix ``Boo1.''

\section{Velocity and Metallicity Measurements}
\label{sec:measurements}
\subsection{Radial Velocity Measurements}
\label{sec:rv}
To determine the heliocentric radial velocities $\rm v_{hel}$ of each star, we use the template-fitting code described in \citet{li2017}. Because no velocity standard stars were observed with VLT, and the Keck/DEIMOS spectra have a much wider wavelength coverage and a similar resolution ($R\sim6,000)$ as the VLT spectra, we use a list of Keck/DEIMOS templates from \citet{kir2015} \revise{that are shifted to zero velocity based on their known velocities. These stellar templates have various effective temperatures, surface gravities, and metallicities, and are fit to the observed spectra using a Markov Chain Monte Carlo (MCMC) sampler \citep{for2013}. }
We adopt a uniform radial velocity prior between $\pm$800 km s$^{-1}$ and use a 100-iteration burn-in to initialize our sampler. The radial velocity posterior distribution is sampled using the ensemble sampler \code{emcee} with 20 walkers and 900 iterations. We then use the median and standard deviation of the posterior chain to compute the radial velocity and radial velocity error, respectively, for each template. We select the radial velocity and error values corresponding to the template fit with the lowest chi-squared value. The radial velocity uncertainties are underestimated and adjusted accordingly (see \S\ref{sec:error}).

We apply several quality checks to our radial velocity results. We use a Random Forest classifier (RFC), as in \citet{li2019}, to identify spectra that do not correspond to stars or good fits. Because the calculated radial velocity of non-stellar or low quality spectra are often large$-$near the upper and lower limits of the allowed $\pm$800 km s$^{-1}$ range$-$we classify any star with an absolute radial velocity greater than 500 km s$^{-1}$ as a poor fit. We then train the RFC using the reduced chi-squared, S/N, absolute deviation, velocity skew, kurtosis and uncertainty of each velocity measurement as features. We apply our trained RFC to the VLT data, providing us with the probability that each star is a good stellar spectrum. We classify all observations with a probability less than 0.6 as bad fits and create a corresponding binary good\_star flag that is set to one for good fits and zero for bad fits.

While the RFC identifies some poor-fitting stellar templates, it does not find all of them. For this reason, we also discard objects with S/N $<4$ or radial velocity uncertainty greater than 20 km s$^{-1}$. After applying these criteria, we have $>$1,000 combined stellar observations, including 113 for Bo{\"o}tes I, 95 for Leo IV and 90 for Leo V. Measurements on these stars are reported in Appendix \ref{sec:allstars}.

We apply the aforementioned measurements to both the combined spectra as well as the spectra from individual exposures. The former are used for membership determination (\S\ref{sec:members}) as well as the velocity dispersion calculation (\S\ref{sec:dispersions}); the latter are used for velocity error correction (\S\ref{sec:error}) and binary search (\S\ref{sec:binaries}).

\subsection{Error Correction}
\label{sec:error}
\begin{figure}
\centering
\includegraphics[width=\columnwidth]{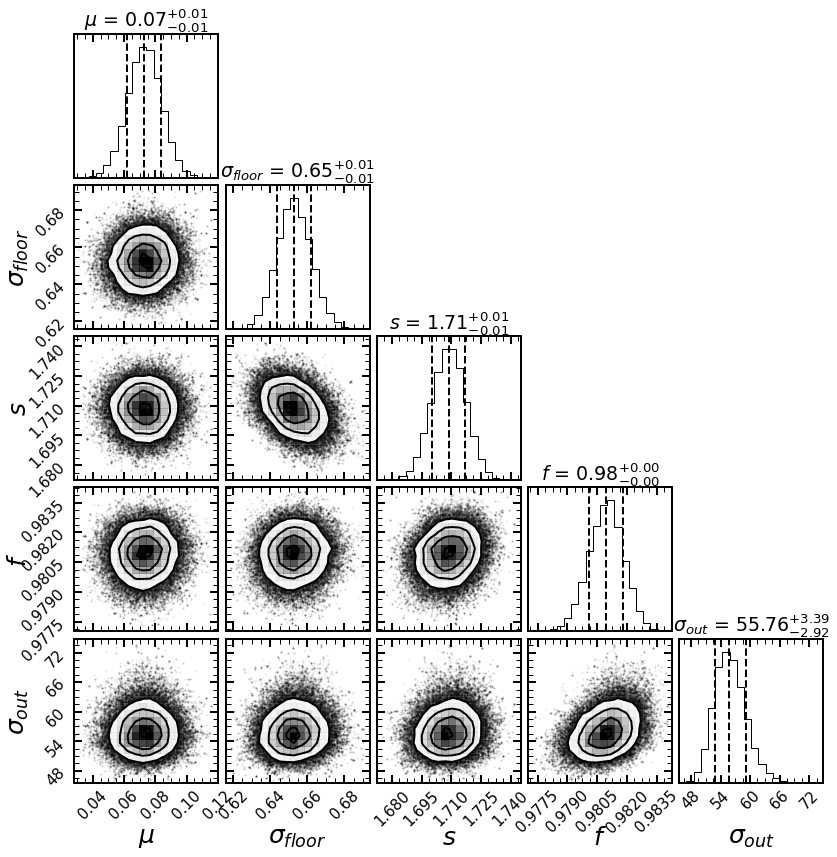}
\caption{Two-dimensional posterior probability distribution from a MCMC sampler using a five-parameter likelihood for radial velocity uncertainty correction. We use the following parameters: the mean pair-wise radial velocity difference $\mu$, the systematic uncertainty floor $\sigma_{floor}$, the uncertainty multiplicative constant $s$, the fraction of non-outliers $f$ and the outlier standard deviation of pair-wise radial velocity differences $\sigma_{out}$. \label{corner_uncertainty}}
\end{figure}
\begin{figure}
\centering
\includegraphics[width=\columnwidth]{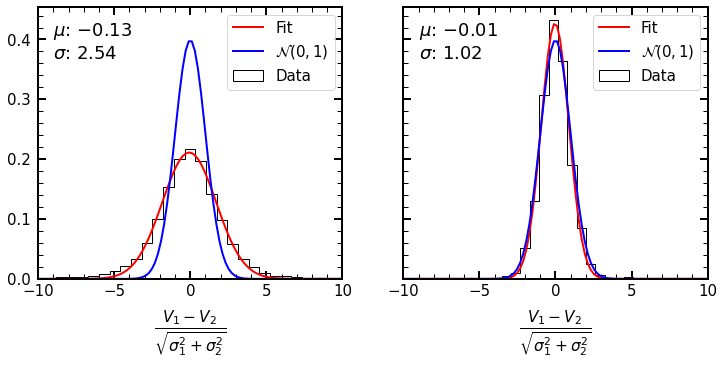}
\caption{Distribution of 40,165 pairwise radial velocity differences divided by the corresponding corrected uncertainties added in quadrature. The red curve depicts a Gaussian fit to the data and the blue curve depicts a standard normal distribution $\mathcal{N}(0,1)$. The means and standard deviations of the data (computed after applying a five sigma clip) are provided in the top left. Left: Pairwise observations from eight UFDs (with more than five observations each) before correction. Right: Pairwise observations after correction, which approximate a normal distribution $\mathcal{N}(0,1)$. \label{error_correction}}
\end{figure}
 
The weighted standard deviation of individual epoch velocities is larger than the corresponding combined radial velocity uncertainty.  Therefore, we use the repeat observations to assess the accuracy of our determined radial velocity uncertainties.
Following \citet{li2019}, we use a Gaussian mixture model to model the pair-wise radial velocity differences $\delta_{i,j} = v_i - v_j$ of stars in the eight UFDs (listed in Table~\ref{tab:table_ufd_list}) with more than five observations. We have
\begin{equation}
P(\delta_{i,j}) = fN(\delta_{i,j}|0,\sqrt{F(\sigma_i)^2+F(\sigma_j)^2}) + (1-f)N(0,\sigma_{out})
\end{equation}
where $N$ is the Gaussian distribution, $\sigma_i$ and $\sigma_j$ are the radial velocity uncertainties corresponding to $v_i$ and $v_j$, and $F$ corresponds to the uncertainty correction function $F(\sigma) = \sqrt{\sigma_{floor}^2 + (s\times\sigma)^2}$. We only include spectra with S/N $>$ 4 and stars with a radial velocity standard deviation $<$ 20 km s$^{-1}$. These quality cuts minimize the effect of poor-quality spectra. 

We find the scaling factor $s$ and systematic floor $\sigma_{floor}$ by fitting the model to 3905 radial velocity pairs. We find $s$ = 1.71 and $\sigma_{floor}$ = 0.65 km s$^{-1}$, resulting in a final uncertainty correction function:
\begin{equation}
F(\sigma) = \sqrt{0.65^2 + (1.71\times\sigma)^2}
\end{equation}
The posterior probability distributions from the MCMC sampler are displayed in Figure~\ref{corner_uncertainty}. The systematic floor is likely due to the limited accuracy of the wavelength calibration, while the multiplicative constant is likely due to the covariance between pixels in the reduced spectra, as the default GIRAFFE pipeline interpolates the extracted wavelengths to a fixed grid. We compare the original and recalibrated uncertainties in Figure~\ref{error_correction} by plotting the distribution of pairwise radial velocity differences divided by the combined uncertainty. The distribution for the recalibrated uncertainties is close to a unit normal distribution, validating our error model. We also apply a similar fit to individual UFDs with more than five observations and obtain similar scaling factors.

\subsection{CaT Metallicity Measurements} 
\label{sec:metallicity}
We determine the metallicities using the calcium triplet (CaT) lines at 8400-8700 \AA. Following the method described by \citet{li2019}, we fit each CaT line with a Gaussian plus Lorentzian function. The resulting sum provides us with an equivalent width (EW). \revise{The EW uncertainty is found from the uncertainty of the fit. In addition, we added a systematic floor of 0.1 \AA in quadrature. This was found using the method described in \S\ref{sec:error}.} The EW is then converted to $\rm[Fe/H]$ using the calibration relation described by \citet{car2013}. The metallicity uncertainties are \revise{propagated from} the CaT EW uncertainties, the photometric uncertainties, distance uncertainties and the uncertainties on the calibration parameters from \citet{car2013}. 

The calibration relation only applies to red giant stars. Additionally, this approach requires an absolute magnitude measurement, which in turn requires that the distance to the star is known. $\rm[Fe/H]$ calculations are therefore only reliable for UFD member stars. For this reason, we do not provide a metallicity literature comparison for all stars. Instead, we compare the literature and measured metallicity values for Bo{\"o}tes I, Leo IV and Leo V member stars in \S\ref{sec:bootes_literature}, \S\ref{sec:leoiv_literature} and \S\ref{sec:leov_literature}. All metallicities for stars with S/N less than seven are discarded.

\section{Member Selection}
\label{sec:members}

Although the data processing and measurements described in \S\ref{sec:observations} and \S\ref{sec:measurements} are applied to all UFDs in Table~\ref{tab:table_ufd_list}, we only perform membership selection for Bo{\"o}tes I, Leo IV and Leo V. We will present membership results for the remaining UFDs in a forthcoming paper. We perform membership classification using both subjective evaluation and a mixture model. The subjective classification is used for subsequent velocity and metallicity calculations in \S\ref{sec:discussion}.

\subsection{Subjective Membership Classification}
We use radial velocity, position, proper motion, metallicity and color-magnitude diagram (CMD) data to evaluate each star's membership subjectively. Location data for each UFD is shown in Figure~\ref{locations}, where we highlight the half-light radii $r_h$ and $\times$3 half-light radii in black, and mark new and previously identified member stars in green and blue, respectively. We expect member stars to lie close to the center of the galaxies. CMDs using dereddened photometry from DECaLS DR8 are presented in Figure~\ref{CMDs}. 
We use a metal-poor ($\feh = -2.3$) Dartmouth isochrone for identifying red giant branch (RGB) candidates and the M92 blue horizontal branch (BHB) ridgeline for identifying BHB candidates. 

Radial velocity and metallicity data are shown in Figure~\ref{metallicities}. We expect member stars to have a radial velocity within $\pm$30 km s$^{-1}$ of the UFD's velocity and to have a lower metallicity characteristic of older stars. We therefore classify all stars with inconsistent radial velocities as non-members and refer to non-member stars that have velocities consistent with the UFD as velocity-consistent non-members (VCNMs). We cross-match the VLT observations with the {\it Gaia} EDR3 catalogue to obtain proper motion and parallax measurements, presented in Figure~\ref{PMs}. We expect member stars to have proper motions consistent with the UFD's motion within three standard deviations. Additionally, we exclude foreground stars using a parallax $\varpi$ cut: $\varpi - 3\sigma_{\varpi}>0$. For stars with uncertain membership, we visually inspect the Mg I line at 8806.8 $\rm\AA$ to determine if the star is a foreground star \citep{bat2011}. This is discussed in Appendix \ref{sec:mg_line}. 

Using these criteria, we identify member stars and VCNMs in the three UFDs, listed in Tables~\ref{tab:table_Bootes_members}, \ref{tab:table_leoiv_members} and \ref{tab:table_leov_members}. Finally, we visually inspect the spectra and confirm that all member stars have qualified measurements on velocity and metallicity fit.  

\subsection{Membership Classification via Mixture Model}
\label{sec:mixturemodel}
We primarily use the above subjective membership selection in \S\ref{sec:discussion}. In addition, we also compute a membership probability to consider how the Milky Way (MW) foreground properties compare to the UFD and whether the exclusion of this affects our results. This probability is calculated by applying a mixture model to each data set.  The mixture model can help justify including or excluding stars on the boundary between the UFD and MW.
We use the spatial position, proper motions from {\it Gaia} EDR3, line-of-sight velocities and metallicities to compute the membership probability.

\begin{figure} 
\centering
\includegraphics[width=\columnwidth]{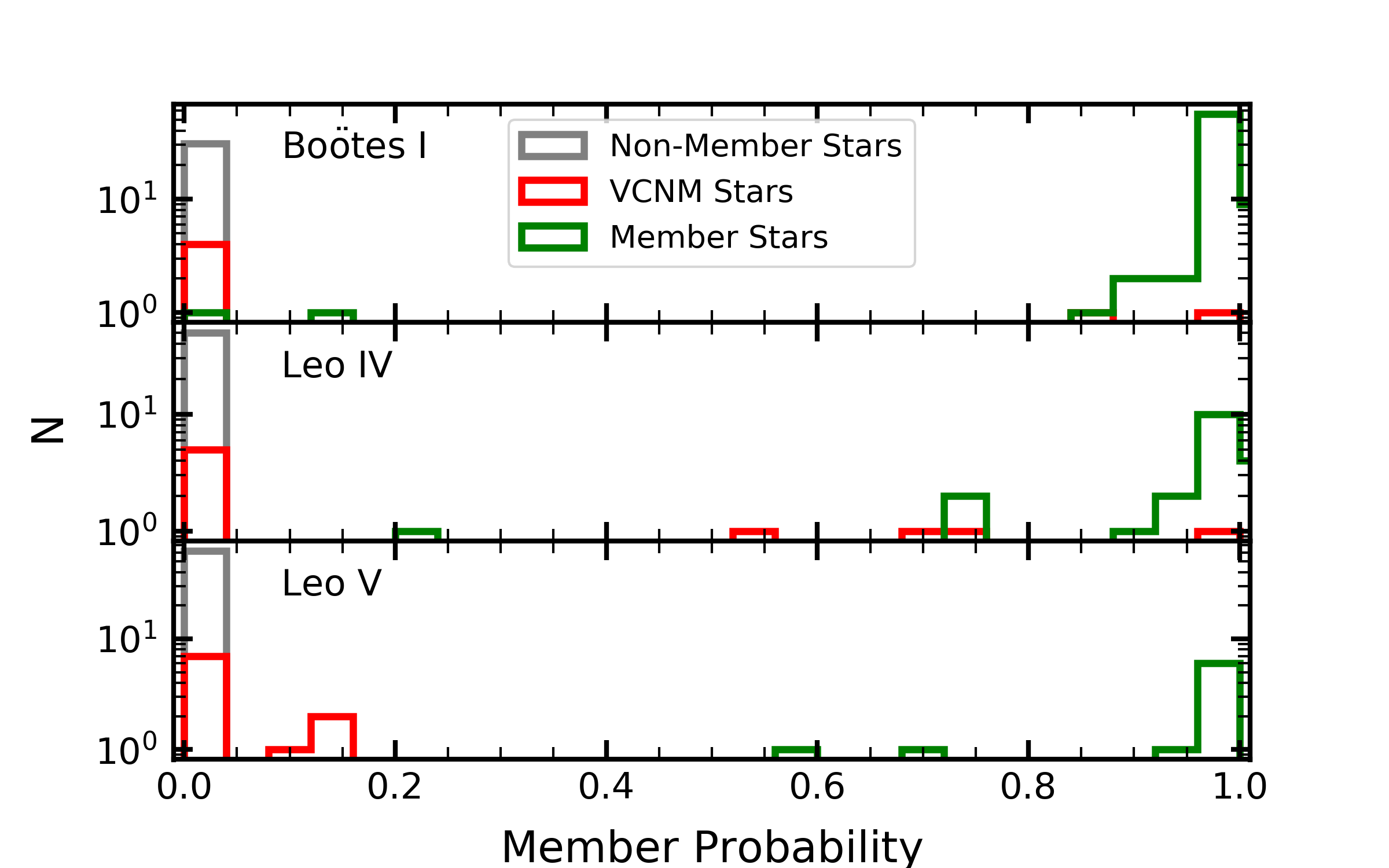}
\caption{ Distribution of Bo{\"o}tes I, Leo IV and Leo V membership probabilities (described in \S\ref{sec:members}), color-coded according to subjective classification. The membership probabilities and subjective classification are largely consistent with each other. \label{Pm}}
\end{figure}

We model the likelihood with a conditional likelihood to account for the unknown spectroscopic selection function \citep[e.g.,][]{Martinez2011ApJ...738...55M, Horigome2020MNRAS.499.3320H}:
\begin{equation}
\mathcal{P}( D| R)  = f(R) \mathcal{P}_{{\rm UFD}}(D|r) + \\(1-f (R)) \mathcal{P}_{\rm MW} (D|R)
\end{equation}
\noindent with $D=\{ v_\mathrm{hel}, \mu_{\alpha \star}, \mu_{\delta}, \feh\}$ and $f(R) = \Sigma_{\rm UFD}(R)/(\Sigma_{\rm UFD}(R) + \Sigma_{\rm MW}(R))$. $\Sigma$ is a 2D density profile.
Here $v_\mathrm{hel}$ is the radial velocity, and $\mu_{\alpha \star}$ and $\mu_{\delta}$ are the proper motions.
The radial density profile is modeled as a Plummer distribution \citep{Plummer1911MNRAS..71..460P}, with best-fit parameters being the projected half-light radii $r_h$, ellipticity $\epsilon$, and position angle $\theta$.  For the spatial parameters ($r_h$, $\epsilon$, $\theta$) we assume Gaussian priors based on the deeper photometric results from \citet{mun2018}.  We fix the center to the values in \citet{mun2018}.
The MW spatial profile is assumed to be constant within the field-of-view of the UFD. 
We model the velocity and metallicity distributions as Gaussian distributions and the proper motion with a multi-variate Gaussian distribution to account for the correlation between the proper motion terms.
As the expected velocity dispersion for a UFD is of order $\sim 5\kms$, which is much smaller than the precision of {\it Gaia} EDR3 proper motions, we only consider a proper motion dispersion for the MW component.
Overall, there are nine parameters to describe the UFD component: $\overline{v}_\mathrm{hel}$, $\sigma_v$, $\overline{{\rm [Fe/H]}}$, $\sigma_{\rm [Fe/H]}$, $\overline{\mu_{\alpha \star}}$, $\overline{\mu_{\delta}}$, $r_h$, $\epsilon$, $\theta$.
There are eight parameters to describe the MW component: $\overline{v}_\mathrm{hel}$, $\sigma_v$, $\overline{{\rm [Fe/H]}}$, $\sigma_{\rm [Fe/H]}$, $\overline{\mu_{\alpha \star}}$, $\sigma_{\mu_{\alpha \star}}$, $\overline{\mu_{\delta}}$, and $\sigma_{\mu_{\delta}}$.  There is one additional parameter: the relative normalization between the UFD and MW spatial distribution.
For the dispersion parameters (e.g., $\sigma_v$, $\sigma_{[Fe/H]}$) we assume log uniform priors (also known as Jeffreys priors). The UFD spatial parameters have Gaussian priors and the remaining parameters have uniform priors.
To compute membership, we compare the ratio of the UFD likelihood to total likelihood: 
\begin{equation}
\begin{aligned}
P(\mathrm{UFD}|D) = \\
\text{\footnotesize $\frac{f(R) P(D|\mathrm{UFD})}{f(R) P(D|\mathrm{UFD}) + (1-f (R)) P(D|\mathrm{MW})}$} 
\end{aligned}
\end{equation}
To compute the posterior distribution we use the \code{MultiNest} algorithm \citep{Feroz2008MNRAS.384..449F,Feroz2009MNRAS.398.1601F}. 
The membership probability ($P_m$) distributions for each UFD are provided in Figure~\ref{Pm}, color coded for different subjectively classified groups: members, non-members and VCNMs. Most members have $P_m > 0.8$ and non-members have $P_m < 0.1$; VCNMs have both high and low membership probabilities. In the following subsections, we discuss in more detail the membership in each UFD and the classification of VCNMs. All measurements, membership probabilities, and subjective membership classifications are given in Appendix \ref{sec:allstars}. Mean velocities and metallicities, velocity and metallicity dispersions, and proper motions calculated using the mixture model are listed in Table~\ref{tab:table_dispersions}.

\subsection{Bo{\"o}tes I Members}
\label{sec:bootes_members}
\begin{figure*}[!htbp]
\centering
\includegraphics[width=2\columnwidth]{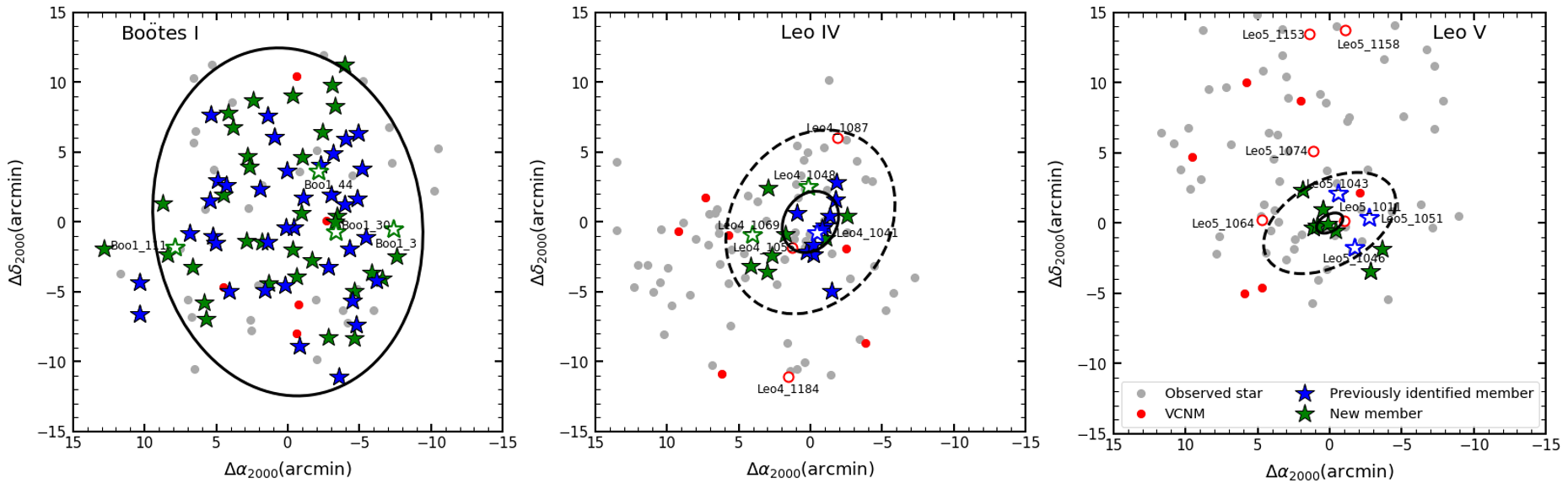}
\caption{Positional data for Bo{\"o}tes I, Leo IV and Leo V, with $r_h$ and $\times$3 $r_h$ shown in black solid and dotted lines, respectively. Blue stars represent previously identified members while green stars represent new members. Red dots represent velocity-consistent non-members (VCNMs). Dark gray points represent stars with a radial velocity more than 30 km s$^{-1}$ from the UFD's systematic velocity. Open symbols are associated with labels and used for identifying specific stars (see main text for details). The same definitions are used in Figures~\ref{CMDs}, \ref{metallicities}, and \ref{PMs}. \label{locations}}
\end{figure*}

\begin{figure*}[!htbp]
\centering
\includegraphics[width=2\columnwidth]{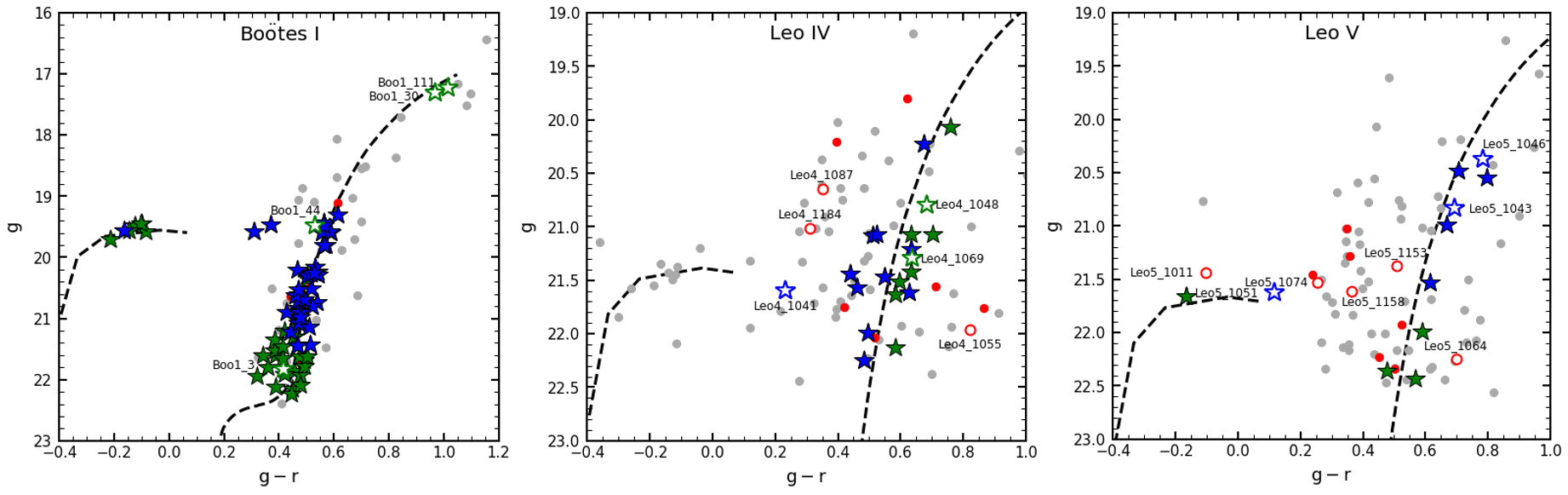}
\caption{CMD for Bo{\"o}tes I, Leo IV and Leo V, with black dotted lines indicating the M92 BHB ridgeline \citep{bel2007} and the Dartmouth isochrone with age = 12.5 Gyr and $\feh = -2.3$ \citep{dot2008}. See Figure~\ref{locations} for definitions of each symbol.   \label{CMDs}}
\end{figure*}

\begin{figure*}[!htbp]
\centering
\includegraphics[width=2\columnwidth]{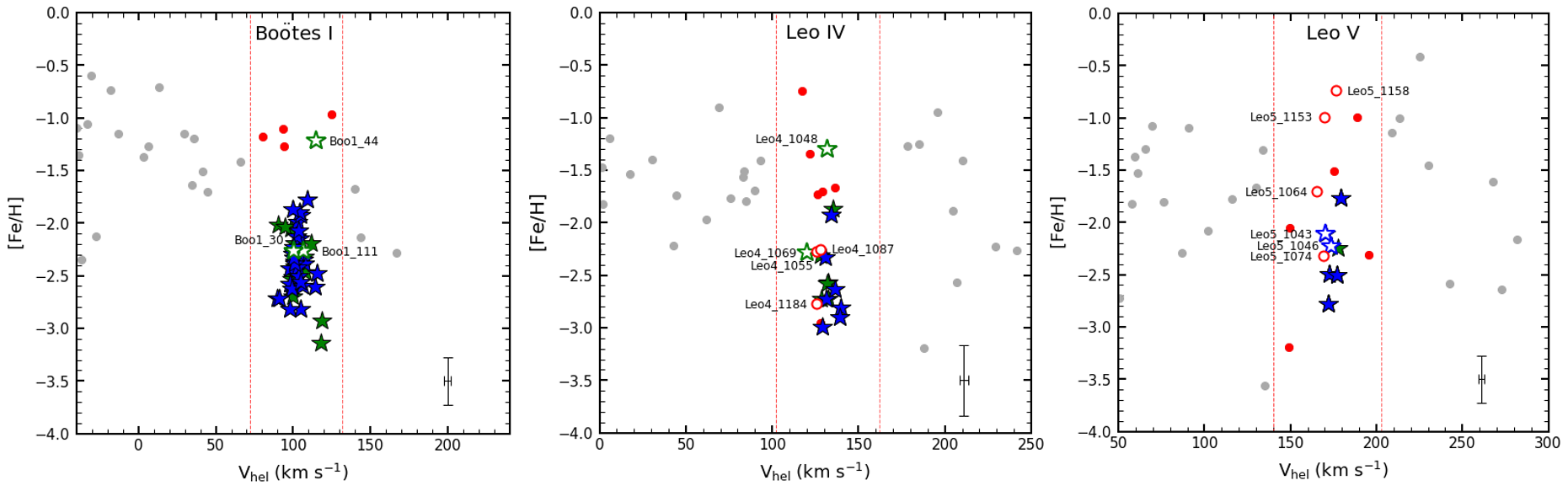}
\caption{Metallicities and radial velocities for Bo{\"o}tes I, Leo IV and Leo V. We only include stars with reliable measurements for both parameters and therefore do not include BHB or RR Lyrae stars. Red dotted lines indicate the upper and lower bounds placed on radial velocity, defined as 30 km s$^{-1}$ above and below the UFD's systematic velocity. \revise{The mean metallicity and radial velocity errors are shown in the bottom right corner of each plot.} See Figure~\ref{locations} for definitions of each symbol.  \label{metallicities}}
\end{figure*}

\begin{figure*}[!htbp]
\centering
\includegraphics[width=2\columnwidth]{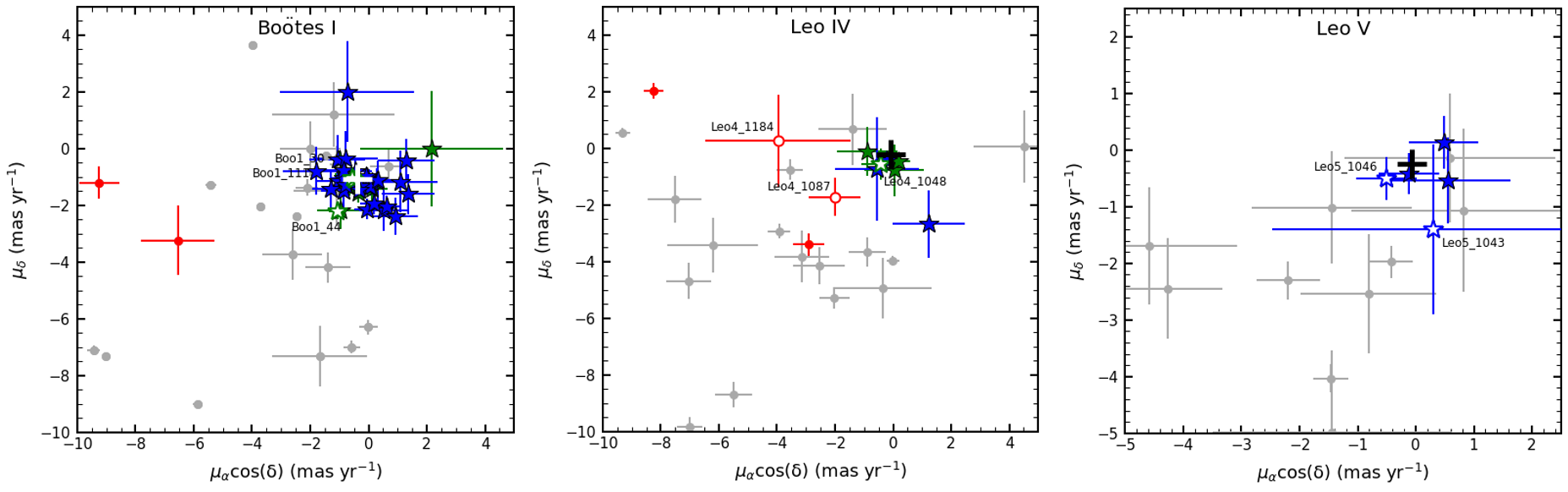}
\caption{{\it Gaia} EDR3 proper motions for Bo{\"o}tes I, Leo IV and Leo V. Crosses indicate UFD proper motions from \citet{mcc2020}. See Figure~\ref{locations} for definitions of each symbol.  \label{PMs}}
\end{figure*}

Using subjective vetting, we identify 69 member stars. Two member stars are assigned a membership probability less than 0.1: Boo1\_3 and Boo1\_44. Boo1\_3's low probability can likely be attributed to its low radial velocity (77.6 $\pm$ 6.0 km s$^{-1}$) while Boo1\_44's low probability can be attributed to its high metallicity ($-$1.22 $\pm$ 0.08 dex). Both stars are highlighted in Figures~\ref{locations}, \ref{CMDs}, \ref{metallicities} and \ref{PMs}. Because their characteristics are otherwise consistent with being member stars, we subjectively classify them as members. All remaining members have a membership probability greater than 0.8. Additionally, all high-probability stars ($P_m >$ 0.5) are classified as members or VCNMs. Two members, Boo1\_30 and Boo1\_111, are located high on the RGB. One member, Boo1\_32, was previously identified as a c-type RR Lyrae star with a period of 0.3119 days by \citet{dal2006}, and is therefore excluded from velocity and metallicity dispersions calculations in Section \ref{sec:dispersions}.

We observe a lack of member stars between $g$ of 17.5 and 19, as seen in Figure~\ref{CMDs}. 
\revise{This is likely because Bo{\"o}tes I was simulatneously observed with both the GIRAFFE spectrograph and Ultraviolet Visual Echelle Spectrograph (UVES) \citep{kop2011}. The bright members at $17.5<g<19$ were targeted by UVES, while here we only present the targets observed with GIRAFFE.}
The lack of bright member stars in the VLT data likely contributes to the proper motion discrepancy discussed in \S\ref{sec:bootes_literature}. 

A velocity histogram of all available radial velocities, shown in Figure~\ref{rv_hists}, has a peak at v $\approx$ 100 km s$^{-1}$. The Leo IV and Leo V velocity distributions are also shown in Figure~\ref{rv_hists}, and metallicity distributions are provided in Figure~\ref{metallicity_distr}.

\subsection{Leo IV Members}
\label{sec:leoiv_members}

We identify twenty Leo IV member stars, including nine new members and eleven members previously identified by \citet{sim2007}. Our subjective membership selection is largely consistent with our membership probabilities. Only one star, Leo4\_1069, is subjectively classified as a member star but assigned membership probability less than 0.5. This is likely due to its lower radial velocity (120.1 $\pm$ 2.8 km s$^{-1}$). Another member, Leo4\_1048, is assigned a membership probability of 1.00, but with large uncertainties. This can be attributed to Leo4\_1048 having a high metallicity ($-$1.30$\pm$0.15 dex) while having a location and proper motion consistent with being a member star. Additionally, Leo4\_1055 is assigned a membership probability of 0.99 but is not classified as a member star because its color-magnitude information is inconsistent with being a Leo IV member.\footnote{However,  \citet{kop2018} found a highly carbon-enhanced extremely metal-poor star in Hydrus I that appears to be much redder than the rest Hydrus I members. Therefore, it could possibly be a member star of Leo IV with an unusual chemical abundance pattern.} As CMD information is not considered when calculating the membership probability, this inconsistency was not accounted for. Two additional stars with consistent velocities and low metallicities, Leo4\_1087 and Leo4\_1184, are considered as VCNMs due to inconsistent proper motions and location on the CMD. \citet{sim2007} had previously classified Leo4\_1087 as a member star. 

We measure the velocity of a previously identified RR Lyrae star: Leo4\_1041. \citet{mor2009} identify Leo4\_1041 (called V2 by \citet{mor2009} and HiTS113256-003329 by \citet{med2018}) as an ab-type RR Lyrae with a period of 0.7096 days.

A velocity histogram of all available velocities (Figure~\ref{rv_hists}) has a peak at v $\approx$ 130 km s$^{-1}$. 
The characteristics of all member stars and VCNMs are given in Table~\ref{tab:table_leoiv_members}.

\subsection{Leo V Members}
\label{sec:leov_members}
We identify eleven Leo V member stars, including four new members, five members previously identified by \citet{wal2009} and \citet{col2017}, one RR Lyrae star identified by \citet{med2017} and one tentative member identified by \citet{mut2020}. One new member star, Leo5\_1014, is a BHB star. The remaining three new members are RGB stars. Two of the previously identified members (Leo5\_1153 and Leo5\_1158) are assigned a membership probability less than 0.2. Leo5\_1153 and Leo5\_1158 were observed by \citet{wal2009} and have low membership probabilities because they are far away ($\sim$13 arcmin, $>$10 $r_h$) from the center of the galaxy.
\citet{mut2020} do not consider these two stars to be members, as the spectra did not pass their quality control cuts and there was no additional information identifying them as members. In this study, we successfully recover their radial velocities and find that they are consistent with Leo V's velocity, suggesting that they could be member stars. If true, this would indicate that Leo V has an extended stellar distribution and support the argument for tidal disruption. However, their metallicity, location and distance from the isochrone are inconsistent with membership, and we therefore do not consider them member stars for subsequent kinematic measurements. 

Additionally, one star with a consistent radial velocity and low metallicity, Leo5\_1074, is not considered a member due to its distance from the Leo V center. Leo5\_1064 similarly has a consistent radial velocity, and is considered a VCNM due to its distance from the Leo V center, distance from the isochrone, and relatively high metallicity. Another star, Leo5\_1011, is a BHB star based on its location on the CMD.  We classify it as a non-member because we measure its velocity to be 25.1$\pm$11.0 km s$^{-1}$ below Leo V's systemic velocity, and no metallicity or proper motion information is available. Furthermore, it is $\sim 0.2$ mag brighter than the other BHB member star. Follow-up observations are needed to determine its membership.

We measure a previously identified RR Lyrae star: Leo5\_1051. \citet{med2017} identify a total of three RR Lyrae stars in Leo V, including Leo5\_1051 (called HiTS113057+021330 by \citet{med2017}), which they classify as an ab-type RR Lyrae with a period of 0.6453 days. Additionally, we identify three stars that \citet{mut2020} listed as plausible Leo V members requiring further follow-up: Leo5\_1046, Leo5\_1043 and Leo5\_1095 (stars 2p, 4p and 3p in \citet{mut2020} Table 10, respectively). Leo5\_1046 was previously observed by \citet{wal2009} but did not pass the quality cuts used by \citet{mut2020}. Leo5\_1046's radial velocity, proper motion, location, metallicity and color-magnitudes are all consistent with Leo V's characteristics, and we classify it as a member star. Leo5\_1043 was identified as a plausible member star because its proper motion and Megacam color-magnitudes were consistent with Leo V. With our spectroscopic data, we confirm that its radial velocity is also consistent, making it another new member star. We find that the final plausible member, Leo5\_1095, is not a member, as its radial velocity differs from that of Leo V by almost 200 km s$^{-1}$. 

\begin{figure}
\centering
\includegraphics[width=\columnwidth]{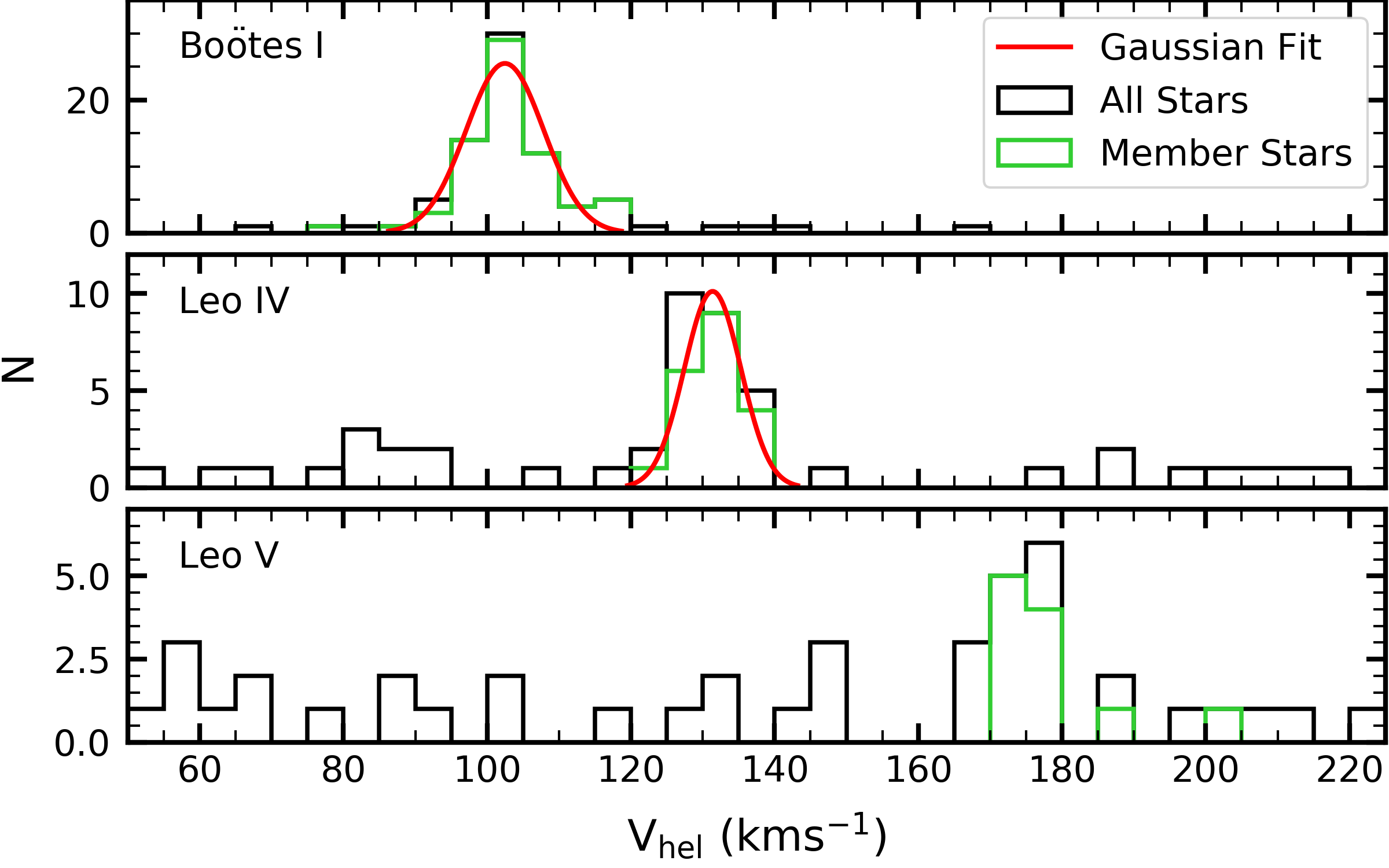}
\caption{Distributions of all available radial velocities (black histograms) and member star radial velocities (green histograms) for Bo{\"o}tes I, Leo IV and Leo V. Red curves depict a re-scaled Gaussian distribution centered at the derived mean velocity with a standard deviation of the velocity dispersion and median uncertainty added in quadrature. Velocity calculations are described in \S\ref{sec:dispersions}. These distributions appear consistent with the distribution of identified members. We note that the Leo V velocity dispersion is not resolved. \label{rv_hists}}
\end{figure}

A velocity histogram of all available velocities (Figure~\ref{rv_hists}) has a peak at v $\approx$ 170 km s$^{-1}$. 
The characteristics of all possible members are given in Table~\ref{tab:table_leov_members}. 
\begin{table*}[t]
\centering
\tiny
\caption{Properties of Bo{\"o}tes I member stars and VCMNs (stars with radial velocities within $\pm$30 km s$^{-1}$ of the UFD's velocity). Column (1) is the star ID, the coordinates are given in columns (2) and (3), and column (4) is the r-band magnitude. Columns (5) and (6) are the measured radial velocities and metallicities, respectively. Column (7) provides the membership probabilities described in \S\ref{sec:members}. Uncertainties on the membership probabilities are reported when the uncertainties are larger than 0.01. Column (8) provides the results of our subjective membership selection, including members (indicated by `M') and VCNMs. The final column contains additional notes. Columns (7) to (9) correspond to columns (8) to (10) in Table~\ref{tab:table_Bootes_members} and Table~\ref{tab:table_leov_members}.} \label{tab:table_Bootes_members}
\centering
\begin{tabular}{lllrrrlcl}
\hline
\hline
 ID& RA & Dec & r & $v_\mathrm{hel}$ & [Fe/H] & $P_m$ & Member & Comments\\
\hline
Boo1\_2 & 209.88958 & 14.47267 & 21.44 & 118.0  $\pm$ 5.2 & $-$3.14  $\pm$ 0.31 & 0.90$^{+0.03}_{-0.04}$ & M & \\ 
Boo1\_3 & 209.89321 & 14.50475 & 21.41 & 77.6  $\pm$ 6.0 & $--$  & 0.14$^{+0.10}_{-0.06}$ & M & \\ 
Boo1\_6 & 209.90667 & 14.44656 & 21.05 & 102.6  $\pm$ 4.0 & $-$2.44  $\pm$ 0.34 & 1.00 & M & \\ 
Boo1\_7 & 209.91400 & 14.44400 & 19.73 & 101.2  $\pm$ 1.1 & $-$2.38  $\pm$ 0.11 & 1.00 & M & \\ 
Boo1\_11 & 209.92575 & 14.49506 & 18.91 & 90.9  $\pm$ 0.7 & $-$2.71  $\pm$ 0.08 & 1.00 & M & \\ 
Boo1\_13 & 209.93100 & 14.57731 & 20.16 & 100.5  $\pm$ 1.6 & $-$2.28  $\pm$ 0.15 & 1.00 & M & \\ 
Boo1\_14 & 209.93575 & 14.61933 & 19.75 & 97.8  $\pm$ 1.1 & $-$2.43  $\pm$ 0.20 & 1.00 & M & \\ 
Boo1\_16 & 209.93729 & 14.54169 & 20.20 & 104.0  $\pm$ 1.5 & $-$1.90  $\pm$ 0.18 & 0.99 & M & \\ 
Boo1\_17 & 209.93775 & 14.39092 & 20.62 & 102.7  $\pm$ 2.8 & $-$1.99  $\pm$ 0.27 & 0.99 & M & \\ 
Boo1\_18 & 209.94046 & 14.43125 & 21.70 & 98.9  $\pm$ 7.1 & $--$  & 0.98 & M & \\ 
Boo1\_19 & 209.94050 & 14.37525 & 21.14 & 90.3  $\pm$ 3.7 & $-$2.02  $\pm$ 0.38 & 0.93$^{+0.03}_{-0.04}$ & M & \\ 
Boo1\_21 & 209.94308 & 14.41994 & 19.75 & 99.3  $\pm$ 1.1 & $-$2.62  $\pm$ 0.10 & 1.00 & M & \\ 
Boo1\_22 & 209.94608 & 14.48125 & 20.98 & 103.5  $\pm$ 3.6 & $-$2.07  $\pm$ 0.55 & 1.00 & M & \\ 
Boo1\_24 & 209.95058 & 14.61294 & 20.20 & 104.4  $\pm$ 1.8 & $-$2.14  $\pm$ 0.26 & 1.00 & M & \\ 
Boo1\_25 & 209.95142 & 14.53431 & 19.27 & 106.8  $\pm$ 0.9 & $-$2.32  $\pm$ 0.08 & 1.00 & M & \\ 
Boo1\_26 & 209.95225 & 14.70122 & 19.69 & 111.5  $\pm$ 1.3 & $-$2.19  $\pm$ 0.12 & 1.00 & M & Binary star\\ 
Boo1\_28 & 209.95888 & 14.32900 & 19.63 & 102.5  $\pm$ 1.9 & $-$2.13  $\pm$ 0.16 & 1.00 & M & \\ 
Boo1\_29 & 209.96146 & 14.52061 & 21.12 & 94.7  $\pm$ 4.5 & $-$2.04  $\pm$ 0.29 & 1.00 & M & \\ 
Boo1\_30 & 209.96212 & 14.50075 & 16.33 & 100.3  $\pm$ 0.7 & $-$2.26  $\pm$ 0.07 & 1.00 & M & \\ 
Boo1\_31 & 209.96283 & 14.51383 & 21.48 & 104.7  $\pm$ 5.2 & $--$  & 0.99 & M & \\ 
Boo1\_32 & 209.96388 & 14.65161 & 19.55 & 113.9  $\pm$ 2.2 & $--$  & 1.00 & M & RR Lyrae star\\ 
Boo1\_33 & 209.96546 & 14.59536 & 18.69 & 100.2  $\pm$ 0.7 & $-$2.33  $\pm$ 0.07 & 1.00 & M & \\ 
Boo1\_36 & 209.96804 & 14.54600 & 20.06 & 100.0  $\pm$ 1.9 & $-$1.87  $\pm$ 0.18 & 1.00 & M & \\ 
Boo1\_37 & 209.97088 & 14.37558 & 21.79 & 102.0  $\pm$ 7.8 & $--$  & 0.98 & M & \\ 
Boo1\_38 & 209.97133 & 14.45953 & 20.23 & 104.9  $\pm$ 1.5 & $-$2.56  $\pm$ 0.15 & 1.00 & M & \\ 
Boo1\_39 & 209.97396 & 14.51553 & 19.96 & 124.7  $\pm$ 1.6 & $-$0.97  $\pm$ 0.15 & 0.00 & VCNM & Inconsistent proper motion, high metallicity\\ 
Boo1\_42 & 209.97871 & 14.62092 & 19.55 & 115.0  $\pm$ 2.7 & $--$  & 0.85$^{+0.08}_{-0.14}$ & M & BHB star\\ 
Boo1\_43 & 209.98054 & 14.58133 & 19.00 & 100.0  $\pm$ 0.8 & $-$2.29  $\pm$ 0.08 & 1.00 & M & \\ 
Boo1\_44 & 209.98325 & 14.57378 & 18.95 & 114.8  $\pm$ 0.8 & $-$1.22  $\pm$ 0.08 & 0.00 & M & \\ 
Boo1\_48 & 209.99104 & 14.46736 & 19.85 & 100.9  $\pm$ 1.1 & $-$2.41  $\pm$ 0.15 & 1.00 & M & \\ 
Boo1\_51 & 210.00100 & 14.54303 & 20.00 & 109.5  $\pm$ 1.4 & $-$1.78  $\pm$ 0.15 & 0.99$^{+0.00}_{-0.01}$ & M & \\ 
Boo1\_52 & 210.00312 & 14.59139 & 19.91 & 119.5  $\pm$ 4.1 & $--$  & 1.00 & M & BHB star\\ 
Boo1\_53 & 210.00412 & 14.52408 & 21.29 & 107.8  $\pm$ 4.3 & $-$2.47  $\pm$ 0.42 & 1.00 & M & \\ 
Boo1\_55 & 210.00638 & 14.36506 & 19.24 & 103.8  $\pm$ 0.8 & $-$2.50  $\pm$ 0.09 & 1.00 & M & \\ 
Boo1\_56 & 210.00692 & 14.41522 & 18.49 & 93.8  $\pm$ 0.7 & $-$1.27  $\pm$ 0.07 & 0.00 & VCNM & Inconsistent proper motion, high metallicity\\ 
Boo1\_57 & 210.00929 & 14.68728 & 20.19 & 80.4  $\pm$ 2.2 & $-$1.18  $\pm$ 0.18 & 0.00 & VCNM & Inconsistent proper motion, high metallicity\\ 
Boo1\_58 & 210.00954 & 14.44817 & 21.41 & 109.0  $\pm$ 5.2 & $--$  & 0.99 & M & \\ 
Boo1\_59 & 210.01017 & 14.38031 & 20.56 & 103.9  $\pm$ 2.6 & $-$2.19  $\pm$ 0.36 & 0.85$^{+0.07}_{-0.15}$ & VCNM & Inconsistent proper motion\\ 
Boo1\_60 & 210.01279 & 14.50656 & 20.91 & 104.8  $\pm$ 2.6 & $-$1.92  $\pm$ 0.30 & 1.00 & M & \\ 
Boo1\_61 & 210.01383 & 14.48094 & 20.13 & 107.2  $\pm$ 1.3 & $-$2.32  $\pm$ 0.34 & 1.00 & M & Binary star\\ 
Boo1\_62 & 210.01450 & 14.66450 & 21.18 & 105.5  $\pm$ 5.4 & $--$  & 0.98 & M & \\ 
Boo1\_63 & 210.02154 & 14.57439 & 20.36 & 104.7  $\pm$ 1.8 & $-$2.82  $\pm$ 0.27 & 1.00 & M & \\ 
Boo1\_64 & 210.02221 & 14.50647 & 20.50 & 100.6  $\pm$ 1.9 & $-$2.23  $\pm$ 0.22 & 1.00 & M & \\ 
Boo1\_65 & 210.02338 & 14.43856 & 19.79 & 105.9  $\pm$ 1.1 & $-$2.60  $\pm$ 0.11 & 1.00 & M & \\ 
Boo1\_66 & 210.03612 & 14.61508 & 20.28 & 102.9  $\pm$ 2.2 & $-$2.32  $\pm$ 0.16 & 1.00 & M & \\ 
Boo1\_67 & 210.04296 & 14.44072 & 21.45 & 101.9  $\pm$ 5.5 & $--$  & 0.99 & M & \\ 
Boo1\_68 & 210.04421 & 14.63994 & 19.22 & 101.1  $\pm$ 0.9 & $-$2.37  $\pm$ 0.09 & 1.00 & M & \\ 
Boo1\_69 & 210.04454 & 14.49011 & 18.99 & 103.9  $\pm$ 0.8 & $-$2.28  $\pm$ 0.08 & 1.00 & M & \\ 
Boo1\_70 & 210.04804 & 14.43222 & 20.97 & 115.6  $\pm$ 3.9 & $-$2.48  $\pm$ 0.93 & 0.98$^{+0.01}_{-0.01}$ & M & \\ 
Boo1\_71 & 210.05096 & 14.48944 & 20.79 & 99.2  $\pm$ 2.6 & $-$2.53  $\pm$ 0.33 & 1.00 & M & \\ 
Boo1\_73 & 210.05383 & 14.55325 & 18.87 & 101.8  $\pm$ 0.7 & $-$2.29  $\pm$ 0.08 & 1.00 & M & \\ 
Boo1\_75 & 210.06112 & 14.65853 & 20.85 & 101.0  $\pm$ 3.2 & $-$2.19  $\pm$ 0.37 & 1.00 & M & \\ 
Boo1\_78 & 210.06588 & 14.57967 & 21.22 & 118.5  $\pm$ 6.2 & $-$2.93  $\pm$ 0.38 & 0.96$^{+0.01}_{-0.02}$ & M & \\ 
Boo1\_80 & 210.06912 & 14.59167 & 19.66 & 100.5  $\pm$ 2.3 & $--$  & 1.00 & M & BHB star\\ 
Boo1\_81 & 210.06925 & 14.49042 & 20.74 & 99.6  $\pm$ 2.5 & $-$2.51  $\pm$ 0.22 & 1.00 & M & \\ 
Boo1\_82 & 210.08579 & 14.62611 & 21.62 & 104.3  $\pm$ 6.3 & $--$  & 0.99 & M & \\ 
Boo1\_84 & 210.09100 & 14.43147 & 20.42 & 89.5  $\pm$ 2.2 & $-$2.72  $\pm$ 0.23 & 1.00 & M & \\ 
Boo1\_85 & 210.09212 & 14.64394 & 19.60 & 99.3  $\pm$ 2.3 & $--$  & 1.00 & M & BHB star\\ 
Boo1\_86 & 210.09350 & 14.55747 & 19.09 & 98.6  $\pm$ 0.8 & $-$2.44  $\pm$ 0.08 & 1.00 & M & \\ 
Boo1\_87 & 210.09721 & 14.43556 & 21.27 & 93.4  $\pm$ 4.5 & $-$1.11  $\pm$ 0.27 & 0.00 & VCNM & High metallicity\\ 
Boo1\_88 & 210.09742 & 14.54589 & 21.13 & 97.7  $\pm$ 6.0 & $--$  & 0.99 & M & \\ 
Boo1\_89 & 210.10483 & 14.56297 & 20.41 & 99.3  $\pm$ 1.6 & $-$2.59  $\pm$ 0.18 & 1.00 & M & \\ 
Boo1\_90 & 210.10625 & 14.48814 & 20.47 & 103.1  $\pm$ 2.2 & $-$2.42  $\pm$ 0.14 & 1.00 & M & \\ 
Boo1\_93 & 210.11075 & 14.49689 & 20.60 & 107.4  $\pm$ 1.6 & $-$2.38  $\pm$ 0.23 & 1.00 & M & \\ 
Boo1\_95 & 210.11267 & 14.64175 & 19.72 & 107.7  $\pm$ 3.3 & $--$  & 1.00 & M & BHB star\\ 
Boo1\_96 & 210.11367 & 14.53875 & 19.73 & 106.5  $\pm$ 1.0 & $-$2.35  $\pm$ 0.23 & 1.00 & M & \\ 
Boo1\_98 & 210.11833 & 14.39794 & 20.96 & 100.0  $\pm$ 3.9 & $-$2.70  $\pm$ 0.27 & 0.98 & M & \\ 
Boo1\_100 & 210.12058 & 14.41728 & 21.24 & 101.3  $\pm$ 5.3 & $-$2.33  $\pm$ 0.32 & 1.00 & M & \\ 
Boo1\_108 & 210.13779 & 14.49992 & 19.03 & 97.6  $\pm$ 0.8 & $-$2.81  $\pm$ 0.09 & 1.00 & M & \\ 
Boo1\_111 & 210.15579 & 14.48278 & 16.22 & 106.9  $\pm$ 0.7 & $-$2.26  $\pm$ 0.07 & 1.00 & M & Binary star\\ 
Boo1\_112 & 210.16483 & 14.47439 & 19.71 & 106.3  $\pm$ 3.9 & $--$  & 1.00 & M & BHB star\\ 
Boo1\_114 & 210.19812 & 14.40333 & 20.42 & 114.1  $\pm$ 2.1 & $-$2.60  $\pm$ 0.31 & 1.00 & M & Binary star\\ 
Boo1\_115 & 210.19821 & 14.44172 & 20.77 & 98.2  $\pm$ 2.4 & $-$2.58  $\pm$ 0.21 & 1.00 & M & \\ 
Boo1\_117 & 210.24133 & 14.48167 & 21.26 & 98.3  $\pm$ 4.3 & $-$2.05  $\pm$ 0.37 & 0.91$^{+0.03}_{-0.03}$ & M & \\  
\hline
\end{tabular}
\end{table*}

\begin{table*}[t]
\centering
\tiny
\centering
\caption{Properties of Leo IV member stars and VCMNs. See Table~\ref{tab:table_Bootes_members} for descriptions of columns (1) to (6) and (8) to (10). Column (7) indicates whether the star is a previously identified member (with Mutual = Yes for members that were previously identified). \label{tab:table_leoiv_members}}
\begin{tabular}{lllrrrrlcl}
\hline
\hline
ID & RA & DEC & $r$ & $v_\mathrm{hel}$ & \feh & Mutual & $P_M$ & Member & Comments\\ 
\hline 
Leo4\_1150 & 173.17683 & $-$0.68997 & 19.81 & 126.2  $\pm$ 2.0 & $-$1.73  $\pm$ 0.15 & No & 0.00 & VCNM & Inconsistent proper motion, lies far from isochrone,\\ 
& & & & & & & & & $>$3 $r_h$ from Leo IV center\\
Leo4\_1051 & 173.19758 & $-$0.53753 & 20.44 & 132.3  $\pm$ 2.1 & $-$2.72  $\pm$ 0.28 & No & 1.00 & M & \\ 
Leo4\_1061 & 173.19867 & $-$0.57681 & 21.33 & 145.6  $\pm$ 8.0 & $--$ & No & 0.75$^{+0.08}_{-0.11}$ & VCNM & High $v_{hel}$, large CaT EW (but low S/N so [Fe/H] is not measured) \\ 
Leo4\_1087 & 173.20888 & $-$0.44464 & 20.29 & 128.0  $\pm$ 2.5 & $-$2.25  $\pm$ 0.24 & No & 0.71$^{+0.09}_{-0.12}$ & VCNM & Inconsistent proper motion, lies far from isochrone\\ 
Leo4\_1057 & 173.21038 & $-$0.49783 & 21.49 & 127.8  $\pm$ 4.8 & $--$ & Yes & 0.94$^{+0.02}_{-0.03}$ & M & \\ 
Leo4\_1045 & 173.21108 & $-$0.51897 & 20.55 & 139.4  $\pm$ 2.2 & $-$2.90  $\pm$ 0.11 & Yes & 1.00$^{+0.00}_{-0.03}$ & M & \\ 
Leo4\_1080 & 173.21583 & $-$0.62719 & 20.92 & 139.9  $\pm$ 3.5 & $-$2.81  $\pm$ 0.74 & Yes & 0.72$^{+0.11}_{-0.18}$ & M & \\ 
Leo4\_1039 & 173.21775 & $-$0.53822 & 20.98 & 134.2  $\pm$ 2.9 & $-$1.92  $\pm$ 0.58 & Yes & 0.99 & M & Binary star\\ 
Leo4\_1043 & 173.22304 & $-$0.56425 & 21.55 & 128.8  $\pm$ 6.1 & $--$ & No & 0.98 & M & \\ 
Leo4\_1036 & 173.22329 & $-$0.54897 & 21.11 & 131.7  $\pm$ 6.8 & $--$ & Yes & 0.99 & M & \\ 
Leo4\_1037 & 173.22692 & $-$0.55308 & 20.58 & 136.2  $\pm$ 2.8 & $-$2.63  $\pm$ 0.35 & Yes & 1.00 & M & \\ 
Leo4\_1041 & 173.23258 & $-$0.55825 & 21.36 & 132.4  $\pm$ 9.2 & $--$ & Yes & 0.99 & M & RR Lyrae star\\ 
Leo4\_1046 & 173.23729 & $-$0.57222 & 20.57 & 129.1  $\pm$ 2.1 & $-$2.99  $\pm$ 0.25 & Yes & 1.00 & M & \\ 
Leo4\_1056 & 173.23750 & $-$0.58386 & 21.76 & 125.3  $\pm$ 7.4 & $--$ & Yes & 0.95$^{+0.02}_{-0.03}$ & M & \\ 
Leo4\_1048 & 173.24300 & $-$0.50364 & 20.11 & 131.8  $\pm$ 1.6 & $-$1.30  $\pm$ 0.15 & No & 1.00$^{+0.00}_{-1.00}$ & M & \\ 
Leo4\_1052 & 173.24458 & $-$0.58056 & 19.55 & 131.2  $\pm$ 1.0 & $-$2.72  $\pm$ 0.12 & Yes & 1.00 & M & \\ 
Leo4\_1040 & 173.25583 & $-$0.53419 & 21.00 & 130.8  $\pm$ 3.2 & $-$2.33  $\pm$ 0.24 & Yes & 1.00 & M & \\ 
Leo4\_1055 & 173.26150 & $-$0.57553 & 21.14 & 125.9  $\pm$ 3.7 & $-$2.27  $\pm$ 0.23 & No & 0.99 & VCNM & Lies far from isochrone\\ 
Leo4\_1184 & 173.26617 & $-$0.73044 & 20.70 & 125.7  $\pm$ 4.0 & $-$2.77  $\pm$ 0.20 & No & 0.55$^{+0.24}_{-0.34}$ & VCNM & Lies off isochrone, $>$3 $r_h$ from Leo IV center\\ 
Leo4\_1050 & 173.26996 & $-$0.55944 & 20.37 & 132.2  $\pm$ 1.8 & $-$2.57  $\pm$ 0.16 & No & 1.00 & M & \\ 
Leo4\_1065 & 173.28504 & $-$0.58472 & 21.05 & 132.1  $\pm$ 3.2 & $-$2.58  $\pm$ 0.40 & No & 0.99 & M & \\ 
Leo4\_1006 & 173.28958 & $-$0.50436 & 19.31 & 128.5  $\pm$ 0.9 & $-$2.72  $\pm$ 0.11 & No & 1.00 & M & \\ 
Leo4\_1077 & 173.29129 & $-$0.60494 & 20.92 & 128.2  $\pm$ 3.8 & $-$2.31  $\pm$ 0.68 & No & 0.90$^{+0.04}_{-0.05}$ & M & \\ 
Leo4\_1069 & 173.30796 & $-$0.56075 & 20.66 & 120.1  $\pm$ 2.8 & $-$2.29  $\pm$ 0.25 & No & 0.22$^{+0.42}_{-0.18}$ & M & \\ 
Leo4\_1085 & 173.31025 & $-$0.59758 & 20.79 & 135.3  $\pm$ 3.4 & $-$1.87  $\pm$ 0.62 & No & 0.76$^{+0.08}_{-0.12}$ & M & \\ 
Leo4\_1010 & 173.33546 & $-$0.56042 & 19.18 & 136.3  $\pm$ 1.2 & $-$1.67  $\pm$ 0.11 & No & 0.00 & VCNM & Inconsistent proper motion\\ 
Leo4\_1204 & 173.34379 & $-$0.72672 & 20.85 & 121.8  $\pm$ 5.0 & $-$1.34  $\pm$ 0.34 & No & 0.01$^{+0.02}_{-0.01}$ & VCNM & Lies far from isochrone, $>$3 $r_h$ from Leo IV center\\ 
Leo4\_1104 & 173.36254 & $-$0.51644 & 20.90 & 117.2  $\pm$ 3.8 & $-$0.74  $\pm$ 0.34 & No & 0.00 & VCNM & High metallicity, lies far from isochrone, $>$3 $r_h$ from Leo IV center\\ 
Leo4\_1142 & 173.39346 & $-$0.55628 & 21.52 & 107.4  $\pm$ 6.8 & $--$ & No & 0.00 & VCNM & Low $v_{hel}$, $>$3 $r_h$ from Leo IV center\\ 
\hline
\end{tabular}
\end{table*}

\begin{table*}[t]
\centering
\tiny
\caption{Properties of Leo V member stars and VCMNs. See Table~\ref{tab:table_Bootes_members} for descriptions of columns (1) to (6) and (8) to (10). Column (7) indicates whether the star is a previously identified member (with Mutual = Yes for members that were previously identified). \label{tab:table_leov_members}}
\centering
\begin{tabular}{lllrrrrlcl}
\hline
\hline
 ID & RA & DEC & $r$ & $v_\mathrm{hel}$ & \feh & Mutual & P$_M$ & Member & Comments\\
\hline 
Leo5\_1062 & 172.72554 & 2.18881 & 21.40 & 177.8  $\pm$ 4.4 & $-$2.25  $\pm$ 0.63 & No & 0.59$^{+0.20}_{-0.28}$ & M & \\ 
Leo5\_1069 & 172.73858 & 2.16256 & 20.92 & 177.1  $\pm$ 2.7 & $-$2.50  $\pm$ 0.23 & Yes & 0.72$^{+0.16}_{-0.30}$ & M & \\ 
Leo5\_1051 & 172.73942 & 2.22514 & 21.51 & 200.3  $\pm$ 13.6 & $--$ & No &  $--$  & M & RR Lyrae star, high $v_{hel}$\\ 
Leo5\_1052 & 172.75150 & 2.25575 & 21.22 & 195.5  $\pm$ 4.9 & $-$2.31  $\pm$ 0.70 & No & 0.01$^{+0.04}_{-0.01}$ & VCNM & High $v_{hel}$\\ 
Leo5\_1046 & 172.75692 & 2.19031 & 19.58 & 173.6  $\pm$ 0.9 & $-$2.22  $\pm$ 0.13 & Yes & 1.00 & M & \\ 
Leo5\_1158 & 172.76729 & 2.44900 & 21.25 & 176.5  $\pm$ 4.5 & $-$0.74  $\pm$ 0.19 & No & 0.01 & VCNM &  $>$5 $r_h$ from Leo V center\\ 
Leo5\_1011 & 172.76917 & 2.22183 & 21.54 & 145.8  $\pm$ 10.8 & $--$ & No &  $--$  & VCNM & Low $v_{hel}$, far from BHB ridgeline\\ 
Leo5\_1043 & 172.77571 & 2.25456 & 20.14 & 170.1  $\pm$ 1.4 & $-$2.11  $\pm$ 0.18 & No & 0.97$^{+0.02}_{-0.08}$ & M & \\ 
Leo5\_1036 & 172.77850 & 2.21078 & 21.88 & 187.8  $\pm$ 6.4 & $--$ & No & 0.95$^{+0.03}_{-0.08}$ & M & \\ 
Leo5\_1032 & 172.78796 & 2.21844 & 21.86 & 173.3  $\pm$ 4.9 & $--$ & No & 0.99 & M & \\ 
Leo5\_1038 & 172.79412 & 2.23597 & 19.75 & 179.4  $\pm$ 1.0 & $-$1.77  $\pm$ 0.11 & Yes & 1.00$^{+0.00}_{-0.06}$ & M & Binary star\\ 
Leo5\_1034 & 172.80021 & 2.21656 & 19.77 & 171.8  $\pm$ 0.9 & $-$2.78  $\pm$ 0.10 & Yes & 1.00 & M & Binary star\\ 
Leo5\_1074 & 172.80458 & 2.30533 & 21.27 & 169.1  $\pm$ 5.3 & $-$2.32  $\pm$ 0.61 & No & 0.15$^{+0.15}_{-0.09}$ & VCNM & $>$5 $r_h$ from Leo V center\\ 
Leo5\_1037 & 172.80500 & 2.21433 & 20.32 & 172.7  $\pm$ 1.6 & $-$2.49  $\pm$ 0.24 & Yes & 1.00 & M & \\ 
Leo5\_1153 & 172.80879 & 2.44344 & 20.87 & 169.7  $\pm$ 3.1 & $-$0.99  $\pm$ 0.26 & No & 0.00 & VCNM & $>$5 $r_h$ from Leo V center\\ 
Leo5\_1014 & 172.81762 & 2.25875 & 21.83 & 176.6  $\pm$ 13.5 & $--$ & No &  $--$  & M & BHB star, $>$3 $r_h$ from Leo V center\\ 
Leo5\_1110 & 172.81962 & 2.36503 & 21.78 & 141.5  $\pm$ 9.8 & $--$ & No & 0.00 & VCNM & Low $v_{hel}$, $>$5 $r_h$ from Leo V center\\ 
Leo5\_1086 & 172.86425 & 2.14267 & 20.93 & 175.1  $\pm$ 4.1 & $-$1.51  $\pm$ 0.20 & No & 0.14$^{+0.15}_{-0.09}$ & VCNM & Lies far from isochrone, $>$5 $r_h$ from Leo V center\\ 
Leo5\_1064 & 172.86454 & 2.22314 & 21.55 & 165.4  $\pm$ 5.2 & $-$1.71  $\pm$ 0.24 & No & 0.09$^{+0.15}_{-0.07}$ & VCNM & Lies far from isochrone, $>$5 $r_h$ from Leo V center\\ 
Leo5\_1129 & 172.88267 & 2.38653 & 21.84 & 188.8  $\pm$ 6.9 & $-$0.99  $\pm$ 0.45 & No & 0.00 & VCNM & High metallicity, $>$5 $r_h$ from Leo V center\\ 
Leo5\_1101 & 172.88454 & 2.13583 & 20.68 & 149.5  $\pm$ 3.0 & $-$2.06  $\pm$ 0.33 & No & 0.00 & VCNM & Lies far from isochrone, low $v_{hel}$, $>$5 $r_h$ from Leo V center\\ 
Leo5\_1124 & 172.94508 & 2.29803 & 21.40 & 149.3  $\pm$ 4.2 & $-$3.20  $\pm$ 0.19 & No & 0.00 & VCNM & Low $v_{hel}$, $>$5 $r_h$ from Leo V center\\ 
\hline
\end{tabular}
\end{table*}

\begin{figure}
\centering
\includegraphics[width=\columnwidth]{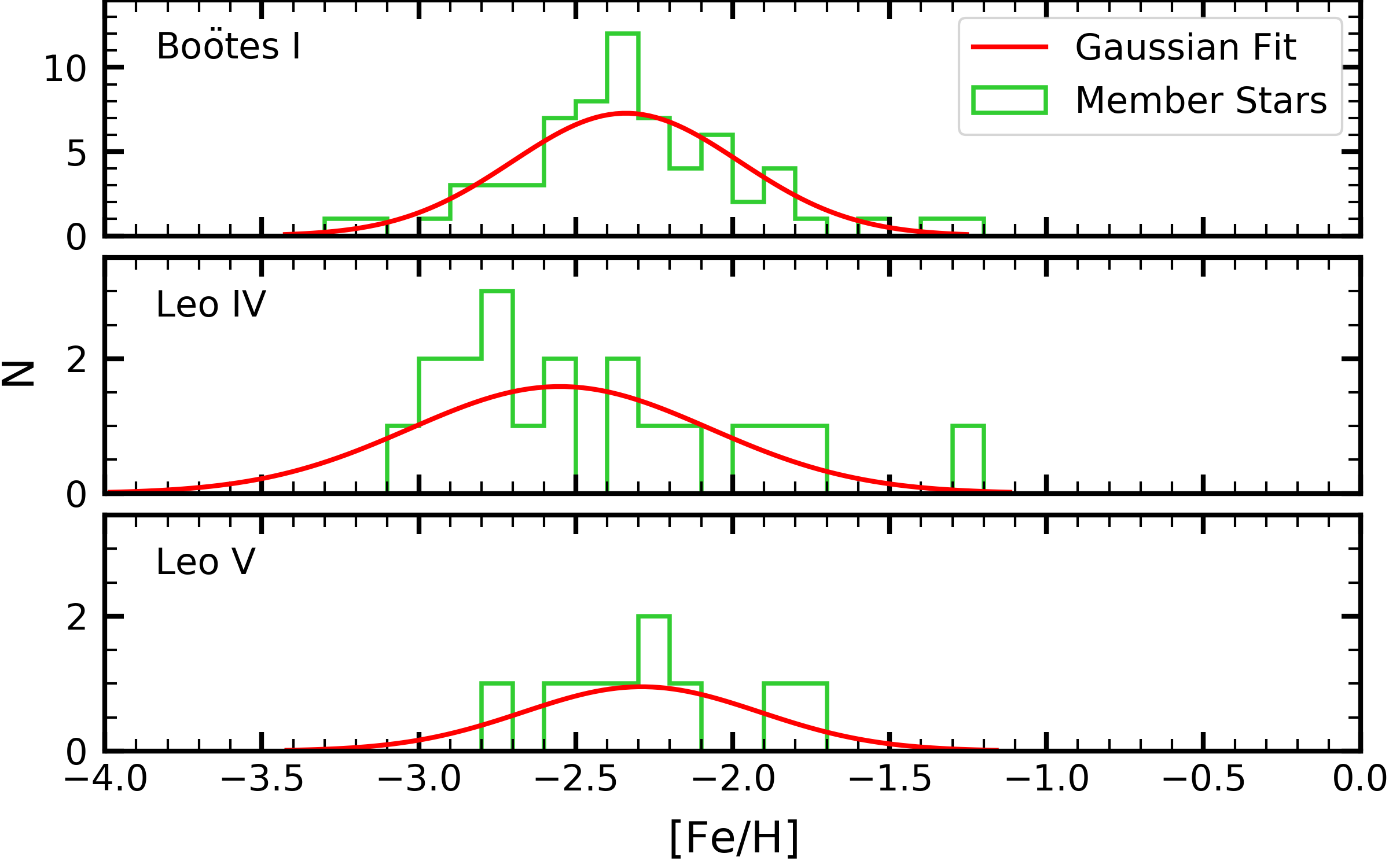}
\caption{ Distributions of all available metallicity measurements for Bo{\"o}tes I, Leo IV and Leo V member stars. Metallicity measurements are described in \S\ref{sec:metallicity} and used in determining membership, as described in \S\ref{sec:members}. Red curves depict a re-scaled Gaussian distribution centered at the derived mean metallicity with a standard deviation of the metallicity dispersion and median uncertainty added in quadrature. Metallicity calculations are described in \S\ref{sec:dispersions}. \label{metallicity_distr}}
\end{figure}
\section{Discussion}
\label{sec:discussion}
\subsection{Binarity} 
\label{sec:binaries}
The measured velocity dispersions could be inflated by the presence of binary systems. To investigate this effect, we measure velocity variation by implementing a $\chi^2$ test on single exposures of the three UFDs. \revise{For each star, we measure the single exposure radial velocities using the best-fit stellar template for the combined spectrum.} We use 21 exposures for Bo{\"o}tes I, 17 exposures for Leo IV and 17 exposures for Leo V (see \S\ref{sec:data}). We perform velocity measurements for single exposures following the same method described in \S\ref{sec:rv}. Using the null hypothesis that the stellar radial velocity is constant, we calculate a p-value for each member star that has a minimum of two exposures with S/N $>$ 4 and radial velocity uncertainty $<$ 20 km s$^{-1}$. For stars with a p-value $<$ 0.1, we visually inspect each exposure to verify the quality of our template-fitting radial velocity measurement. This check allows us to discard outliers that are due to noise rather than variability. We then re-calculate the p-value for such stars. 

Additionally, both Leo IV and Leo V observations were taken in two groups several months apart. We leverage this to identify potential binary stars by co-adding the spectra taken within each time frame and re-calculating the p-value. This method is especially useful for faint stars that have low S/N from single-exposure spectra.

We can confidently reject the null hypothesis (p$<$0.01) for one star in Bo{\"o}tes I: Boo1\_26. Additionally, we find weak evidence (0.01$\le$p$\le$0.1) that three Bo{\"o}tes I member stars exhibit variability: Boo1\_61, Boo1\_111 and Boo1\_114. Three of the four potential binaries (Boo1\_26, Boo1\_61 and Boo1\_111) were assigned a variability probability greater than 0.9 by \citet{kop2011} and were thus considered too variable to include in kinematic measurements. In contrast, \citet{kop2011} assigned a variability probability of 0.61 to Boo1\_114 and therefore did not classify it as a binary. We visually inspect the single-exposure spectra and confirm that there is distinct velocity variation. It is not used in subsequent kinematics calculations. We do not find evidence of variability for two stars that \citet{kop2011} identified as potential binaries: Boo1\_30 and Boo1\_32. However, Boo1\_32 is a previously identified RR Lyrae star and we therefore do not include it in any kinematic calculations. Boo1\_30 exhibits a peak-to-peak radial velocity variation of just 2.5$\pm$1.7 km s$^{-1}$ and is included in kinematic calculations. 

We do not identify any potential binary stars in Leo IV when using single-exposure spectra. However, when we use the two combined spectra, we find that Leo4\_1039 exhibits weak evidence of variability, with a p-value of 0.06. We calculate velocities of 141.84$\pm$5.46 km s$^{-1}$ and 129.20$\pm$4.04 km s$^{-1}$ for the first and second combined observing periods, respectively. This star is too faint to obtain reliable velocity measurements from single-exposure spectra.

\begin{figure*}
\centering
\includegraphics[width=2\columnwidth]{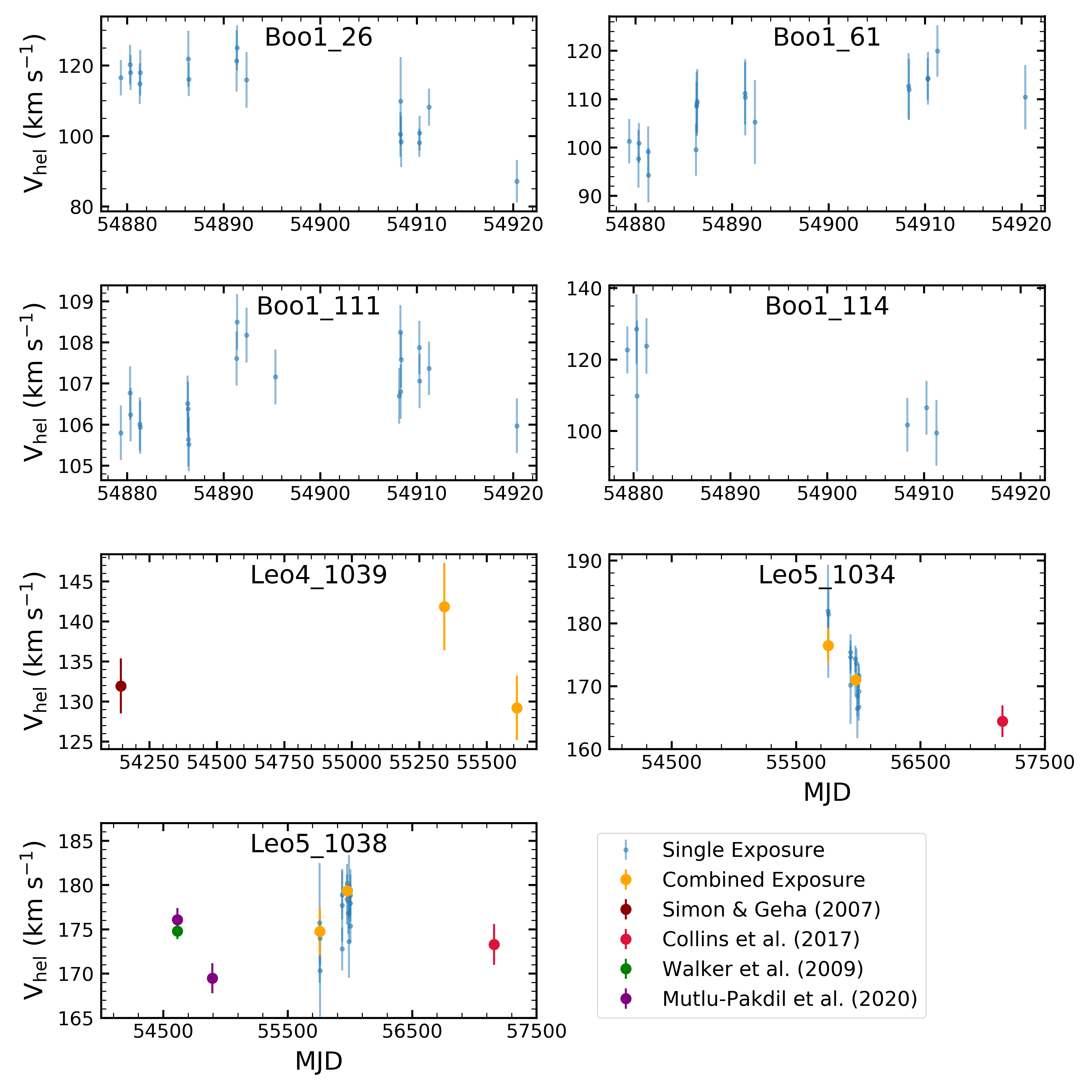}
\caption{ Radial velocities of potential binary stars. We present our single-exposure measurements (blue) and literature radial velocities. Because the Leo IV and Leo V single exposures can be clustered into two groups observed approximately eight and seven months apart, respectively, we separately combine exposures within each observing cluster. The resulting combined radial velocities are shown in orange. Aside from Leo5\_1038, all stars show evidence of velocity variation using the VLT measurements alone. Leo5\_1038 is a binary found in \citet{mut2020} and our radial velocity measurements support this claim. For Leo4\_1039, we are only able to measure its radial velocity from the combined exposures due to the low S/N of single-exposure spectra. \label{1038}
}
\end{figure*}

We identify one star in Leo V with evidence of variability in velocity \revise{(p$=$0.06)} using the combined measurements: Leo5\_1034. After investigating the quality of each exposure, we conclude that Leo5\_1034's variability appears to be due to true fluctuations in velocity. Additionally, \citet{mut2020} identified Leo5\_1038 as a possible binary. Though we calculate a p-value of 0.57 (single exposures) and 0.11 (combined exposures) for this member, our results are consistent with this claim. As seen in Figure~\ref{1038}, where we compare our measurement to all previous measurements, Leo5\_1038's radial velocity appears to change with time. We calculate a weighted standard deviation of 3.22 km s$^{-1}$, 2.2 times larger than the mean uncertainty of 1.45 km s$^{-1}$. 

We note that although we identify RR Lyrae members in both Leo IV and Leo V, the S/N from single-exposure spectra are too low to derive reliable velocity measurements for velocity variation studies. All potential binaries are displayed in Figure~\ref{1038}.

\subsection{Velocity and Metallicity Dispersions} 
\label{sec:dispersions}
Using our expanded membership catalogues, we re-calculate the mean velocities $\overline{v}_\mathrm{hel}$ and velocity dispersions $\sigma_v$ for each UFD. We use the two-parameter log-likelihood function from \citet{wal2009}. $v_i$ and $\sigma_i$ are the measured radial velocity and internal uncertainty, respectively, for the $i$th of N stars. While $\sigma_i$ captures the random internal measurement uncertainty, $\sigma_V$ captures the intrinsic radial velocity dispersion. We assume that the measured radial velocities have a Gaussian distribution centered on $\overline{v}_\mathrm{hel}$. A uniform prior is used for $\overline{v}_\mathrm{hel}$. Both a log uniform prior and uniform prior are used for $\sigma_v$; we report the former as M1 in Table~\ref{tab:table_dispersions} and the latter as M2. M1 is considered the default value.

The two parameters of interest, $\overline{v}_\mathrm{hel}$ and $\sigma_V$, are estimated by sampling the following distribution: 
\begin{equation}\label{eq:dispersion_eq}
\log \mathcal{L}=-\frac{1}{2}\sum_{i=1}^{N}\ln(\sigma_i^2+\sigma_V^2)-\frac{1}{2}\sum_{i=1}^N\frac{(v_i-\overline{v}_\mathrm{hel})^2}{(\sigma_i^2+\sigma_V^2)}.
\end{equation}

We restrict $\sigma_v$ to be greater than 0 km s$^{-1}$ and $\overline{v}_\mathrm{hel}$ to be within the range of member star velocities. We run an MCMC sampler for 5,000 iterations, with a burn-in period of 50 iterations and priors chosen randomly from a normal distribution centered at the member stars' mean velocity. We use a similar approach to determine the mean metallicities $\overline{\rm[Fe/H]}$ and metallicity dispersion $\sigma_{\rm[Fe/H]}$ from our CaT metallicity measurements. 

We calculate the mean velocities $\overline{v}_\mathrm{hel}$, velocity dispersions $\sigma_v$, mean metallicities $\overline{\rm[Fe/H]}$ and metallicity dispersions $\sigma_{\rm[Fe/H]}$ of the three dwarf galaxies using the subjectively identified member stars listed in Tables~\ref{tab:table_Bootes_members}-\ref{tab:table_leov_members}. For kinematic calculations, we remove potential binary stars and RR Lyrae stars identified in \S\ref{sec:members} and \S\ref{sec:binaries}, which may potentially inflate the velocity dispersion if included. For metallicity calculations, only bright RGB members are included as we do not have reliable \feh measurements for horizontal branch stars, nor RGB members with S/N $<$ 7.
This results in a total of 64, 18 and 8 members in Bo{\"o}tes I, Leo IV, Leo V in the velocity dispersion calculations and 51, 15 and 7 members in the metallicity dispersion calculations. 

The velocity and metallicity dispersions calculated using a log uniform  and uniform prior are largely consistent. We do, however, observe a discrepancy in the Leo V velocity dispersion; using a log uniform prior we calculate a dispersion of \revise{0.1$^{+1.5}_{-0.1}$ km s$^{-1}$} and using a uniform prior we calculate a dispersion of \revise{2.4$^{+2.2}_{-1.6}$ km s$^{-1}$}. Due to the small number of stars included in the calculation, the prior does have an impact to the results. In both cases, we conclude that the Leo V velocity dispersion is not fully resolved. However, when we include the binary stars Leo5\_1034 and Leo5\_1038, we resolve the velocity dispersion, calculating a value of 3.0$^{+1.3}_{-1.0}$ km s$^{-1}$. 

As discussed in \S\ref{sec:bootes_members}, Boo1\_3 has a low membership probability of 0.14 due to its low velocity (more than 20 km s$^{-1}$ below the systemic velocity of Bo{\"o}tes I). Due to its faintness, the proper motion of this star is unavailable; moreover, we are unable to reliably measure its metallicity using the spectrum with S/N = 4. Despite these limitations, we classify it as a member star in the subjective membership classification based on its spatial location and location on the CMD. We recalculate the velocity dispersion excluding this star and find a slightly lower velocity dispersion of 4.8$^{+0.6}_{-0.5}$ km s$^{-1}$. This is still consistent with the dispersion when the star is included (\revise{5.1$^{+0.7}_{-0.8}$ km s$^{-1}$}).

We identify a high metallicity member star in both Bo{\"o}tes I and Leo IV (Boo1\_44 and Leo4\_1048 with metallicities of $-$1.22 $\pm$ 0.08 and $-$1.30 $\pm$ 0.15 dex, respectively). To evaluate their impact on the metallicity calculations, we re-calculate the mean metallicities and metallicity dispersions without these stars. When excluding Boo1\_44, we find no significant difference in the Bo{\"o}tes I mean metallicity, but calculate a lower metallicity dispersion (0.18$^{+0.04}_{-0.03}$ dex,  compared to \revise{0.28$^{+0.04}_{-0.03}$ dex} when including Boo1\_44). Additionally, when we exclude Leo4\_1048, we calculate a mean Leo IV metallicity of $-$2.66$^{+0.07}_{-0.07}$ dex (compared to \revise{$-$2.48$^{+0.16}_{-0.13}$ dex} when including Leo4\_1048). Furthermore, we are no longer able to resolve the Leo IV metallicity dispersion (0.02$^{+0.06}_{-0.01}$ dex) after excluding this star.

For comparison, we also list the velocity and metallicity parameters from the membership probability calculation discussed in \S\ref{sec:members}. This approach uses all stars with good radial velocity and metallicity measurements, as no prior membership or binary information is available before the mixture model analysis is performed. Given the minimal difference in the subjective membership classification and membership probability (Figure~\ref{Pm}), these two membership classifications provide similar kinematic and metallicity parameters. 
However, as stated previously, we obtain different values for the Bo{\"o}tes I metallicity dispersion depending on our treatment of Boo1\_44, and we find that our value is only consistent with the dispersion from the membership probability calculation when we exclude this star. This is because the star is assigned a membership probability of zero and so is not considered in the dispersion calculation in the mixture model. 

Additionally, while the Leo IV metallicity dispersions are consistent, the mixture model calculates a large uncertainty. This reflects the large uncertainty in Leo4\_1048's membership probability (1.00$^{+0.00}_{-1.00}$). Because Leo4\_1048 is a member star with a metallicity of $-$1.30$\pm$0.16 dex, its inclusion in metallicity calculations has a large effect on the calculated metallicity dispersion. Just as we are unable to resolve Leo IV's metallicity dispersion when we exclude Leo4\_1048 from our member catalogue, the mixture model is unable to resolve the metallicity dispersion when Leo4\_1048 is assigned a membership probability of zero. 

While we do not resolve the Leo V velocity dispersion, the mixture model calculates a value of 3.2$^{+1.7}_{-1.4}$ km s$^{-1}$. This is consistent with our value when we include the two binary member stars and is a consequence of the mixture model not excluding binary stars from velocity calculations. 

In Figures~\ref{rv_hists} and \ref{metallicity_distr} we show the velocity and metallicity distribution of all member stars in these three UFDs, overplotted with a Gaussian function centered at the derived mean velocity/metallicity with a standard deviation of the velocity/metallicity dispersion and median uncertainty added in quadrature. Table~\ref{tab:table_dispersions} lists these main properties of the three UFDs.

\begin{table}[htp!]
\centering
\notsotiny
\label{tab:obslog}
\centering
\caption{UFD properties from literature are given in the top panel. In the bottom panel, the toal number of member stars identified, used in velocity calculations and used in metallicity calculations are reported, in addition to the updated mean velocities ($\overline{v}_\mathrm{hel}$), velocity dispersions ($\sigma_V$), mean metallicities ($\overline{\rm[Fe/H]}$), metallicity dispersions ($\sigma_{\rm[Fe/H]}$) and proper motion ($\mu_{\alpha}\rm cos(\delta)$, $\mu_{\delta}$). Results are reported for the complete membership catalogue using both log uniform priors (M1) and uniform priors (M2) for dispersion parameters. We consider M1 to be the default values used throughout the study. Additionally, we provide velocity and metallicity values found using the membership probability mixture described in \S\ref{sec:mixturemodel} (M3).} \label{tab:table_dispersions}
\begin{tabular}{lrrrr}
\hline
\hline
 Parameter & & Bootes I & Leo IV & Leo V \\
\hline
RA$_\mathrm{J2000}$ (deg) & &  210.0200 &  173.2405 &  172.7857\\
Dec$_\mathrm{J2000}$ (deg) & & 14.5135 & $-$0.5453 & 2.2194\\
M$_V$ (mag) & & $-$6.02$^{+0.25}_{-0.25}$ & $-$4.99$^{+0.26}_{-0.26}$ & $-$4.29$^{+0.36}_{-0.36}$ \\
r$_h$ (pc) & & 191$^{+8}_{-8}$ & 114$^{+13}_{-13}$ & 49$^{+16}_{-16}$ \\
Distance (kpc) & & 66.0$^{+2.0}_{-2.0}$ & 154.0$^{+5.0}_{-5.0}$ & 169.0$^{+4.0}_{-4.0}$ \\
$\overline{v}_\mathrm{hel}$ (km s$^{-1}$) & & 101.8$^{+0.7}_{-0.7}$ & 132.3$^{+1.4}_{-1.4}$ & 170.9$^{+2.1}_{-1.9}$ \\
$\sigma_{v}$ (km s$^{-1}$) & & 4.6 $^{+0.8}_{-0.6}$ & 3.3$^{+1.7}_{-1.7}$ & 2.3$^{+3.2}_{-1.6}$ \\
$\overline{\rm[Fe/H]}$ (dex)& & $-$2.35$^{+0.09}_{-0.08}$& $-$2.29$^{+0.19}_{-0.22}$ & $-$2.48$^{+0.21}_{-0.21}$\\
$\sigma_{\rm[Fe/H]}$ (dex)&  & 0.44$^{+0.07}_{-0.06}$ & 0.56$^{+0.19}_{-0.14}$ & 0.47$^{+0.23}_{-0.13}$\\
$\mu_{\alpha}\rm cos(\delta)$ (mas yr$^{-1}$) & & $-$0.39$\pm$0.01 & $-$0.08$\pm$0.09 & $-$0.06$\pm$0.09\\
$\mu_{\delta}$ (mas yr$^{-1}$)& & $-$1.06$\pm$0.01 & $-$0.21$\pm$0.08 & $-$0.25$^{+0.09}_{-0.08}$\\
References$^a$ & & 1,1,1,1,2,3, & 1,1,1,1,6,7, & 1,1,1,1,9,10,\\
 & & 3,4,4,5,5 & 7,8,8,5,5 & 10,10,10,5,5\\
\hline
\# of members & Total & 69  & 20 & 11\\
             & RV    & 64 & 18 & 8 \\
             & \feh  & 51 & 15 & 7 \\
$\overline{v}_\mathrm{hel}$ (km s$^{-1}$) & \textbf{M1} & \textbf{102.6$^{+0.7}_{-0.8}$} & \textbf{131.6$^{+1.0}_{-1.2}$}  &\textbf{173.1$^{+1.0}_{-0.8}$} \\
  & M2 & 102.5$^{+0.8}_{-0.7}$ & 131.2$^{+1.1}_{-1.2}$ & 173.5$^{+2.0}_{-1.3}$ \\
  & M3 & 102.9$^{+0.7}_{-0.7}$  & 131.5$^{+1.0}_{-0.9}$ & 174.1$^{+1.7}_{-1.5}$ \\
$\sigma_V$ (km s$^{-1}$) & \textbf{M1} & \textbf{5.1$^{+0.7}_{-0.8}$} & \textbf{3.4$^{+1.3}_{-0.9}$}&\textbf{0.1$^{+1.5}_{-0.1}$} \\
  & M2 & 5.2$^{+0.8}_{-0.7}$ & 3.5$^{+1.4}_{-0.8}$ &  2.3$^{+2.2}_{-1.6}$\\
  & M3 &  5.1$^{+0.6}_{-0.5}$ & 2.7$^{+1.2}_{-1.0}$ & 3.2$^{+1.7}_{-1.4}$ \\
$\overline{\rm[Fe/H]}$ (dex)& \textbf{M1} & \textbf{$-$2.34$^{+0.05}_{-0.05}$} & \textbf{$-$2.48$^{+0.16}_{-0.13}$} & \textbf{$-$2.29$^{+0.14}_{-0.17}$}\\
  & M2 &  $-$2.34$^{+0.05}_{-0.05}$ & $-$2.48$^{+0.18}_{-0.13}$ & $-$2.30$^{+0.17}_{-0.12}$ \\
  & M3 & $-$2.36$^{+0.03}_{-0.03}$  & $-$2.54$^{+0.17}_{-0.15}$ & $-$2.30$^{+0.17}_{-0.17}$ \\
$\sigma_{\rm[Fe/H]}$ (dex)& \textbf{M1} & \textbf{0.28$^{+0.04}_{-0.03}$} & \textbf{0.42$^{+0.12}_{-0.10}$} & \textbf{0.30$^{+0.14}_{-0.09}$}\\
  & M2 &  0.28$^{+0.04}_{-0.04}$ & 0.43$^{+0.15}_{-0.09}$ & 0.34$^{+0.17}_{-0.12}$ \\
  & M3 & 0.14$^{+0.04}_{-0.05}$ & 0.39$^{+0.14}_{-0.33}$ & 0.34$^{+0.16}_{-0.12}$ \\
 $\mu_{\alpha}\rm cos(\delta)$ (mas yr$^{-1}$)& M3 &  $-$0.45$^{+0.04}_{-0.04}$& $-$0.11$^{+0.24}_{-0.24}$& $-$0.02$^{+0.29}_{-0.29}$\\
 $\mu_{\delta}$ (mas yr$^{-1}$)& M3 & $-$1.13$^{+0.03}_{-0.03}$ & $-$0.45$^{+0.19}_{-0.19}$ & $-$0.40$^{+0.21}_{-0.21}$\\
\hline
\end{tabular}
\tablenotetext{a}{References: (1) \citet{mun2018}; (2) \citet{dal2006}; (3) \citet{kop2011}; (4) \citet{sim2019}; (5) \citet{mcc2020}; (6) \citet{mor2009}; (7) \citet{sim2007}; (8) \citet{kir2013}; (9) \citet{med2018}; (10) \citet{col2017}}
\end{table}

\subsection{Mass-to-Light Ratios}
\label{sec:ml_ratio}
With the new spectroscopically confirmed members, we calculate the mass contained within the half-light radius using the estimator from \citet{wol2010}:
\begin{equation}
M_{1/2} = 930\sigma_v^2 r_h M_\odot
\end{equation}
We use the velocity dispersions calculated in \S\ref{sec:dispersions} and the half-light radii measured by \citet{mun2018}. The derived dynamical masses $M_{1/2}$ for Bo{\"o}tes I and Leo IV are $4.9^{+1.3}_{-1.2} \times 10^6$ and $1.3^{+0.8}_{-0.8} \times 10^6$ M$_\odot$, respectively. Because the Leo V velocity dispersion is not resolved, we instead compute the 95\% upper limit. We calculate the upper limit to be $8.9\times 10^5$ M$_\odot$. 

We calculate the luminosity from the V-band magnitudes listed in Table~\ref{tab:table_dispersions}. The  resulting mass-to-light ratios for Bo{\"o}tes I ad Leo IV are $449^{+251}_{-184}$ and $315^{+322}_{-222}$ M$_\odot$/L$_\odot$, respectively. The 95\% upper limit for Leo V is $399$ M$_\odot$/L$_\odot$.
\begin{figure*}
\centering
\includegraphics[width=\textwidth]{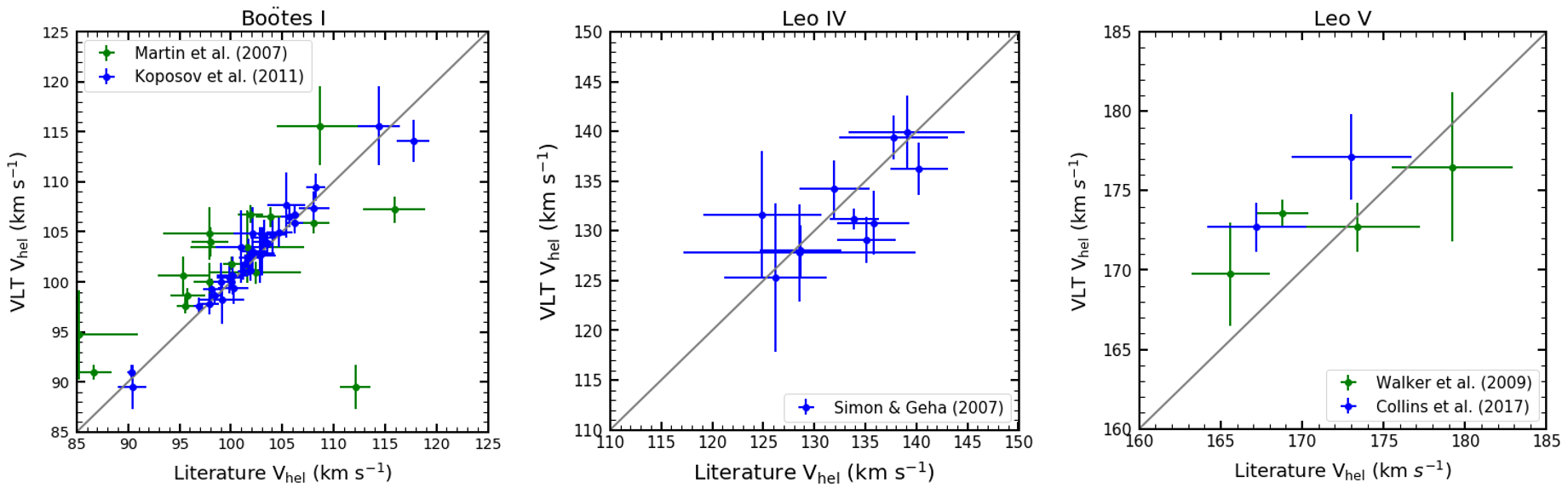}
\caption{Comparison of Bo{\"o}tes I, Leo IV and Leo V member stars' radial velocity measurements with literature values, excluding binary and RR Lyrae stars. Because there is only one Leo V star (Leo5\_1038) in both the VLT data and \citet{mut2020} data with quality spectra, and this star is a binary (see Figure~\ref{1038}), we do not include a comparison with \citet{mut2020}. We do not observe a significant offset for either Bo{\"o}tes I or Leo IV. However, for Leo V, we observe a positive offset of 4.8 $\pm$ 2.9 km s$^{-1}$ from \citet{col2017} based on two common stars. 
\label{rv_comparison}
}
\end{figure*}
\begin{figure*}
\centering
\includegraphics[width=\textwidth]{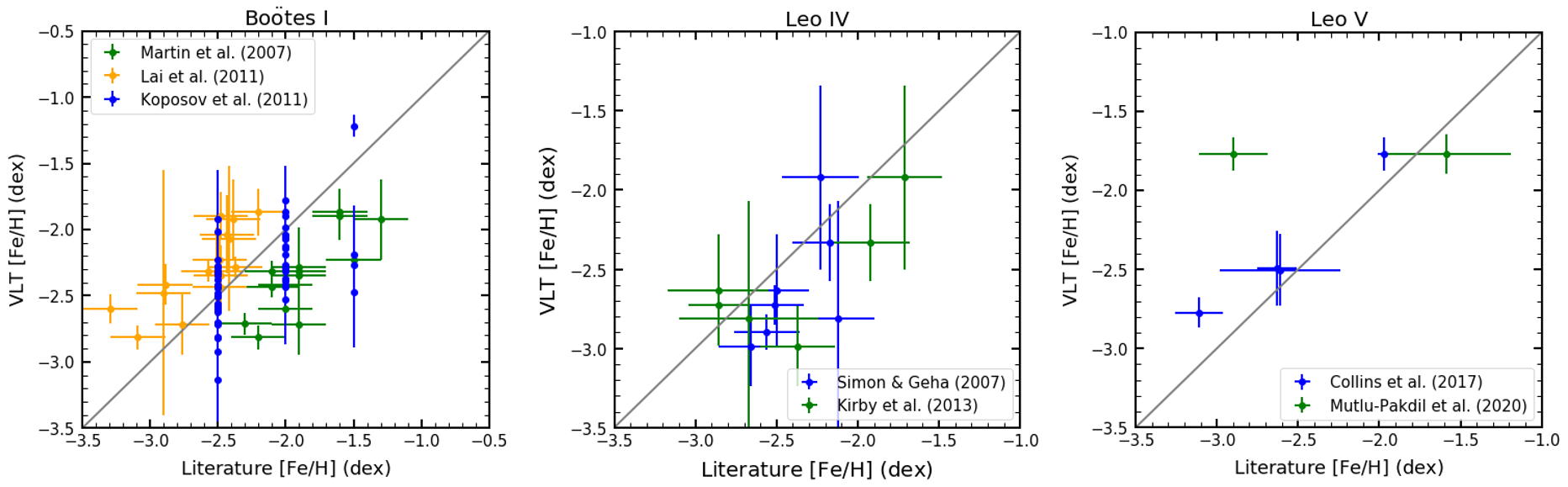}
\caption{Comparison of Bo{\"o}tes I, Leo IV and Leo V member stars' metallicity measurements with literature values. \citet{mut2020} observe the same Leo V member star (Leo5\_1038) twice; we compare our metallicity to both of their measurements, which differ by 1.31$\pm$0.45 dex. This star is considered a binary (see Figure~\ref{1038}). \label{metallicity_comparison}}
\end{figure*}
\subsection{Bo{\"o}tes I Literature Comparison}
\label{sec:bootes_literature}
\citet{kop2011} used an enhanced data reduction pipeline to calculate radial velocities from VLT spectra. They targeted stars that are likely to be members, including nine member stars identified by \citet{mun2006} and 27 member stars identified by \citet{Mar2007}. \citet{kop2011} observed a total of 118 stars, 100 of which were used for kinematic calculations. Though they consider most of the observed stars to be Bo{\"o}tes I members, they classified 37 of them as highly probable members using the following criteria: low velocity variability, log(g) $<$ 3.5, [Fe/H] $<$ $-$1.5 and radial velocity error $dv <$ 2.5 km s$^{-1}$. 

Because we use the data from the same observations analyzed by \citet{kop2011}, we expect to measure similar velocities, metallicities and dispersions. \citet{kop2011} used 100 stars to calculate the velocity dispersion, and identified a subset of 37 stars as highly probable Bo{\"o}tes I members. We assign a membership probability greater than 0.8 to all \citet{kop2011} highly probable members. Of the remaining 36 stars with high membership probability, \citet{kop2011} disqualified 6 because of their variability, 21 for having a radial velocity error $\ge$ 2.5 km s$^{-1}$, 12 for having log(g) $\ge$ 3.5, and 9 for having [Fe/H] $\ge$ $-$1.5 (with 11 stars not meeting at least two of the four criteria). Because the \citet{kop2011} membership cuts are intended to capture highly probable members, they may be excluding less obvious members. The variability and radial velocity error cuts in particular may be broadened to capture more members. 

We calculate a mean velocity and velocity dispersion of \revise{102.6$^{+0.7}_{-0.8}$ and 5.1$^{+0.7}_{-0.8}$ km s$^{-1}$}, respectively. These values are consistent with the \citet{kop2011} values for a single-component distribution ($\overline{v}_\mathrm{hel}$ = 101.8$^{+0.7}_{-0.7}$ km s$^{-1}$, $\sigma_v$ = 4.6$^{+0.8}_{-0.6}$ km s$^{-1}$). We observe velocity offsets of 0.36 $\pm$ 0.37 km and 2.0 $\pm$ 1.0 km s$^{-1}$ from \citet{kop2011} and \citet{Mar2007}, respectively. We compare member stars' velocity measurements to previous literature in the left panel of Figure~\ref{rv_comparison}.

\begin{figure*}
\centering
\includegraphics[width=2\columnwidth]{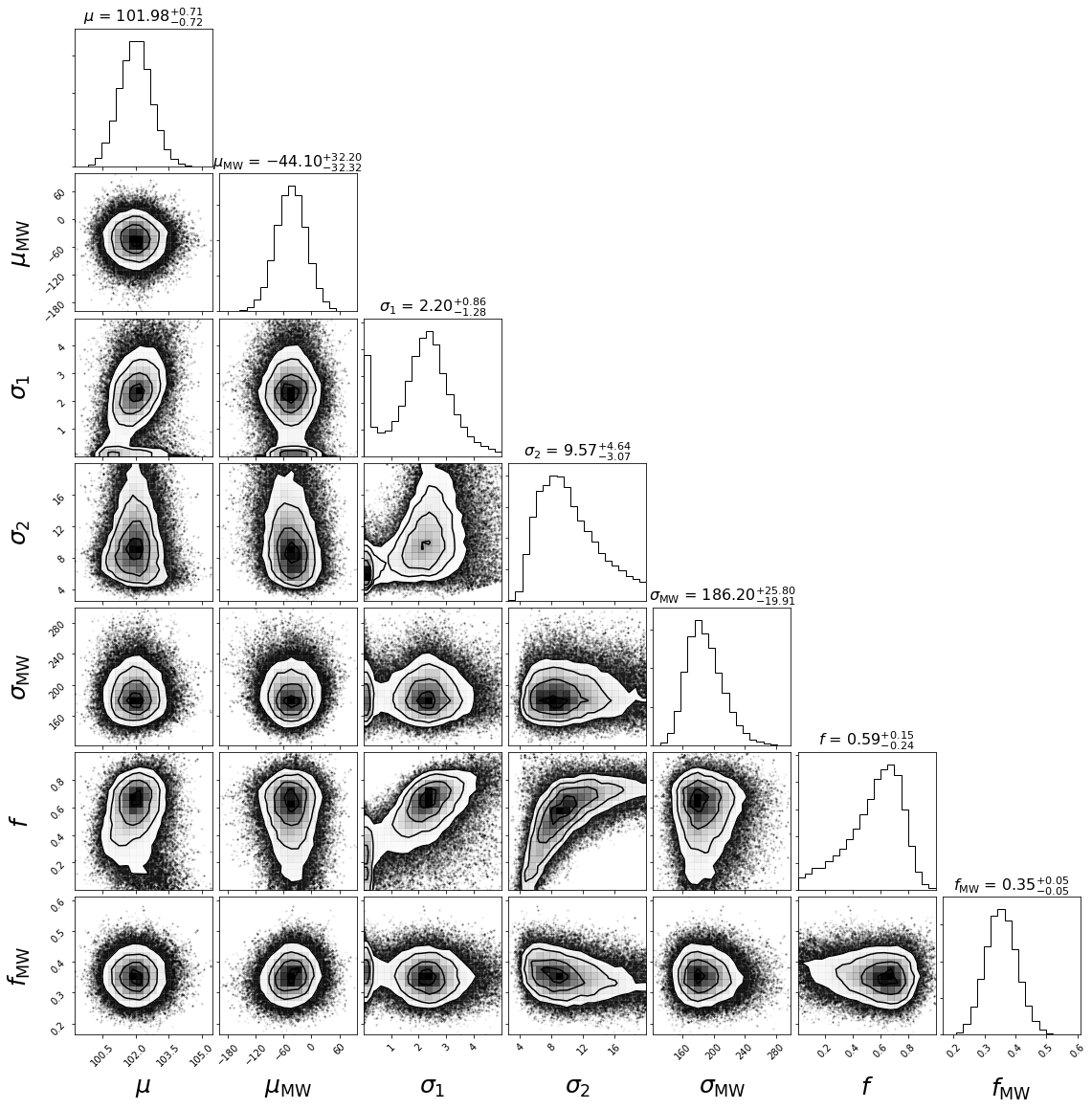}
\caption{ \revise{Two-dimensional posterior probability distribution from a MCMC sampler using a seven-parameter likelihood. We use the following parameters: the mean velocity of Bo{\"o}tes I $\mu$, the mean velocity of the foreground MW stars $\mu_{MW}$, the velocity dispersion of the first Bo{\"o}tes I velocity component $\sigma_1$, the velocity dispersion of the second Bo{\"o}tes I velocity component $\sigma_2$, the velocity dispersion of the MW stars $\sigma_{MW}$, the fraction of Bo{\"o}tes I stars in the first velocity component $f$, and the fraction of MW stars $f_{MW}$.} \label{2comp}
}
\end{figure*}

\citet{kop2011} also fitted a two-component velocity distribution and found a cold component with $\sigma_v\sim2.4 \kms$ and a hot component with $\sigma_v\sim9 \kms$. 
We fit a similar two component model using \revise{a seven parameter likelihood that includes a MW component: 
\begin{equation}
\begin{aligned}
P(v) = (1-f_{MW}) \times (fN(v|\mu,\sigma_1) + \\
(1-f)N(\mu,\sigma_2)) + f_{MW}N(\mu_{MW},\sigma_{MW})
\end{aligned}
\end{equation}
$\mu$ and $\mu_{MW}$ are the mean velocities of Bo{\"o}tes I and MW stars, respectively; $\sigma_1$, $\sigma_2$, and $\sigma_{MW}$ are the velocity dispersions of the first Bo{\"o}tes I velocity component, the second Bo{\"o}tes I velocity component, and the MW stars, respectively; and $f$ and $f_{MW}$ are the fraction of Bo{\"o}tes I stars in the first velocity component and the fraction of stars in the foreground, respectively. Using this likelihood, we calculate velocity dispersions ($\sigma_v\sim2.2 \kms$ and  $\sigma_v\sim9.6 \kms$) consistent with those found by \citet{kop2011}. The posterior probability distribution is shown in Figure \ref{2comp}. We also fit a corresponding one component model using a five parameter likelihood (i.e no $\sigma_2$ and $f$). }

\revise{We compare the models using the corrected Akaike Information Criterion (AICc), which is a likelihood ratio with an additional penalty for the number of model parameters (see \citealt{kir2013} for details). 
A smaller AICc value usually corresponds to a more favored model. Our calculated AICc value for the two-component model is lower than one-component model by 3.5, indicating the two-component model is preferred, but with weak evidence.
} 

We calculate a mean metallicity and metallicity dispersion of \revise{$-$2.34$^{+0.05}_{-0.05}$ and 0.28$^{+0.04}_{-0.03}$ dex}, respectively. The mean metallicity is consistent with the most recent calculation given by \citet{sim2019}. However, our metallicity dispersion is smaller than that from \citet{sim2019} (0.44$^{+0.07}_{-0.06}$ dex). \citet{kop2011} use the best-fit template for each star's co-added spectrum to calculate metallicities. Though their fitting procedure is primarily intended to calculate radial velocities, they find that their resulting values for $T_\mathrm{eff}$, log($g$) and [Fe/H] are reasonable. We compare our metallicity results with theirs in Figure~\ref{metallicity_comparison}, noting that they do not provide uncertainties for their measurements because of the discreteness of the Munari atmosphere grid \citep{mun2005}. Assigning their values a uniform uncertainty of 0.3 dex, we find an offset of $-$0.03 $\pm$ 0.06 dex. There is no obvious correlation between our measurements. We also provide a comparison of our results with \citet{Mar2007} and \cite{lai2011}. Measurements from both studies are known to within $\pm$0.2 dex. We find an offset of $-$0.47$\pm$0.11 and 0.35$\pm$0.11 dex relative to the measurements from \citet{Mar2007} and \citet{lai2011}, respectively. This could be attributed in part to the different CaT calibrations used; while we use the calibration relation from \citet{car2013}, \citet{Mar2007} use the relation from \citet{rut1997}. The comparison results are shown in the left panel of Figure~\ref{metallicity_comparison}.

Using the membership mixture model described in \S\ref{sec:mixturemodel}, we calculate the systematic proper motion in right ascension and declination to be $-$0.45$^{+0.04}_{-0.04}$ and $-$1.13$^{+0.03}_{-0.03}$ mas yr$^{-1}$, respectively. \citet{mcc2020} similarly use the {\it Gaia} EDR3 catalogue to measure the systematic proper motion of Bo{\"o}tes I, calculating the right ascension and declination components to be $-$0.39 $\pm$ 0.01 and $-$1.06 $\pm$ 0.01 mas yr$^{-1}$, respectively. The discrepancy between these measurements is likely due to the lack of bright member stars in the VLT data. This is discussed in more detail in \S\ref{sec:bootes_members}.

\subsection{Leo IV Literature Comparison}
\label{sec:leoiv_literature}
\citet{sim2007} used Keck/DEIMOS spectroscopy to identify eighteen Leo IV members. We recover eleven of these stars and identify nine additional members. We calculate a mean velocity and velocity dispersion of \revise{131.6$^{+1.0}_{-1.2}$ and 3.4$^{+1.3}_{-0.9}$} km s$^{-1}$, respectively. \citet{sim2007} calculated the mean velocity and velocity dispersion using a maximum-likelihood method that assumes a Gaussian velocity distribution, as in \citet{wal2006}. They found values of 132.3 $\pm$ 1.4 km s$^{-1}$ and 3.3 $\pm$ 1.7 km s$^{-1}$, respectively. These are consistent with the values calculated using our member catalogue. The \citet{sim2007} dispersion is consistent with zero within two standard deviations. Our measurement is therefore the first to resolve the Leo IV velocity dispersion at the 95\% confidence level.

The VLT data includes member stars that have previously been identified, providing a source of direct comparison for member velocities. We compare the radial velocity measurements of eleven member stars to previously published findings in the middle panel of Figure~\ref{rv_comparison}. We observe that our VLT data for member stars is on average 0.8 $\pm$ 2.0 km s$^{-1}$ lower than the literature values. All previously identified members are listed in Table~\ref{tab:table_leoiv_previousmembers} in Appendix \ref{sec:mutualstars}, along with all corresponding velocity measurements.

We calculate a mean metallicity and metallicity dispersion of \revise{$-$2.48$^{+0.16}_{-0.13}$ and 0.42$^{+0.12}_{-0.10}$ dex}, respectively. These values are consistent with those calculated by \citet{kir2013}. We compare our calculated metallicities to those measured by \citet{sim2007} and \citet{kir2013} in the middle panel of Figure~\ref{metallicity_comparison}. We find a mean offset of $-$0.22 $\pm$ 0.17 and $-$0.17 $\pm$ 0.21 dex from \citet{sim2007} and \citet{kir2013}, respectively. The offset from \citet{sim2007} is consistent with the difference between the mean metallicities calculated here and by \citet{sim2007} (0.16$\pm$0.17 dex). \revise{This discrepancy can be attributed in part to the different CaT calibrations used. As discussed in \S\ref{sec:bootes_literature}, we use the calibration relation from \citet{car2013}, while measurements prior to 2013 use the relation from \citet{rut1997}.}

Using the membership mixture model, we calculate the systematic proper motion in right ascension and declination to be $-$0.11$^{+0.24}_{-0.24}$ and $-$0.45$^{+0.19}_{-0.19}$ mas yr$^{-1}$, respectively. This is consistent with the values found by \citet{mcc2020}.

\subsection{Leo V Literature Comparison}
\label{sec:leov_literature}
\citet{wal2009} identified a total of seven Leo V members observed with MMT/Hectochelle and \citet{col2017} performed follow-up with Keck/DEIMOS spectra, finding an additional five members. \citet{mut2020} used photometric data and two epochs of stellar spectra observed with MMT/Hectochelle to find three new possible member stars. One of the epochs of spectroscopic data was first reported by \citet{wal2009}. 
We identify eleven member stars, including four new members, two members previously observed in all three studies, three members observed by \citet{wal2009} and two members observed by \citet{col2017}. \citet{wal2009} classified two stars (Leo5\_1153 and Leo5\_1158) located far ($\sim$13 arcmin, $>$10 $r_h$) from the center of Leo V as members. These stars are highlighted in Figures~\ref{locations} through \ref{PMs}. We classify them as VCNMs due to their distance from the Leo V center and their high metallicities.

We calculate a mean velocity of \revise{173.1$^{+1.0}_{-0.8}$ km s$^{-1}$} and are unable to resolve the velocity dispersion.  \citet{wal2009} used a two-dimensional likelihood function weighted by membership probabilities to perform kinematic calculations. Using only the five central members, they found a velocity dispersion of 2.4$^{+2.4}_{-1.4}$ km s$^{-1}$. This value is not conclusively resolved. Including the two distant candidate members increases this value to 3.7$^{+2.3}_{-1.4}$ km s$^{-1}$. \citet{col2017} estimated Leo V's kinematics using two models: one that assumes the system is dispersion supported and one that allows for a velocity gradient. For the former, they calculated the mean velocity to be 172.1$^{+2.3}_{-2.1}$ km s$^{-1}$ and the velocity dispersion to be 4.0$^{+3.3}_{-2.3}$ km s$^{-1}$; for the latter, they calculated the mean velocity to be 170.9$^{+2.1}_{-1.9}$ km s$^{-1}$ and the velocity dispersion to be 2.3$^{+3.2}_{-1.6}$ km s$^{-1}$. Our mean velocity is consistent with the values found by both \citet{wal2009} and \citet{col2017}.

We compare the radial velocity measurements of the seven previously identified member stars to published findings in the right panel of Figure~\ref{rv_comparison}. Our VLT velocity measurements for member stars are on average 4.8 $\pm$ 2.9 km s$^{-1}$ higher than the measurements from \citet{col2017} and 1.4 $\pm$ 2.1 km s$^{-1}$  higher than the measurements from \citet{wal2009}. Our results for all common stars (members and non-members) are more consistent; we observe 3.4 $\pm$ 3.4 km s$^{-1}$ and 0.7 $\pm$ 0.9 km s$^{-1}$ offsets from \citet{col2017} and \citet{wal2009}, respectively, excluding four common stars with a difference $>$ 100 km s$^{-1}$ from our comparison with \citet{wal2009}. We visually verify the spectra quality for all stars with difference $>$ 100 km s$^{-1}$ and do not identify any poor-quality velocity template fits. All large differences are observed relative to \citet{wal2009} and could be attributed to \citet{wal2009} using an older MMT/Hectochelle pipeline that is improved in \citet{mut2020}. We measure an offset of 6.6 $\pm$ 1.8 km s$^{-1}$ from \citet{mut2020} based on the common binary member star. All previously identified members are listed in Table~\ref{tab:table_leov_previousmembers} in Appendix \ref{sec:mutualstars}, along with all corresponding velocity measurements.

We calculate a mean metallicity and metallicity dispersion of \revise{$-$2.29$^{+0.14}_{-0.17}$ and 0.30$^{+0.14}_{-0.09}$ dex}, respectively. These values are consistent with \citet{col2017}. We compare our metallicity results to previous literature (\citet{col2017} and \citet{mut2020}) in the right panel of Figure~\ref{metallicity_comparison}. The two metallicities from \citet{mut2020} in Figure~\ref{metallicity_comparison} are calculated from different spectra for the same star. The variability in their results may suggest that their uncertainty is underestimated. The average offset from \citet{col2017} is 0.19 $\pm$ 0.14 dex.

Using the membership mixture model, we calculate the systematic proper motion in right ascension and declination to be $-$0.02$^{+0.29}_{-0.29}$ and $-$0.40$^{+0.21}_{-0.21}$ mas yr$^{-1}$, respectively. This is consistent with the values found by \citet{mcc2020}.

\subsection{Metallicity Distribution Function of Bo{\"o}tes I}

The VLT data roughly doubles the number of metallicities available in Bo{\"o}tes I compared to the most recent analysis by \citet{lai2011}, bringing it to a total of ${\sim}70$ stars: 51 in the VLT/GIRAFFE data (29 new stars), and 19 stars in the literature not in our dataset \citep{nor2010, lai2011}. We calibrate the literature measurements to our metallicities using the offset calculated in \S\ref{sec:bootes_literature}. 

We re-analyze the metallicity distribution function using three analytic metallicity distribution functions (MDFs) considered by \citet{lai2011}: the leaky box, the pre-enriched leaky box, and the extra gas model \citep{kir2011}.
The leaky box is the classic analytic model characterized by the effective yield $p$.
The pre-enriched box model adds a minimum metallicity floor $\feh_0$.
The extra gas model \citep{lyndenbell75} adds pristine gas to a leaky box parameterized by $M$, where $M=1$ reproduces the leaky box and $M > 1$ adds extra pristine gas to the leaky box, creating a more peaked MDF with a lighter metal-poor tail.
 We note that with 41 stars, \citet{lai2011} found all three models fit the data about equally well, with a slight preference for the extra gas model.

\begin{figure}
    \centering
    \includegraphics[width=\linewidth]{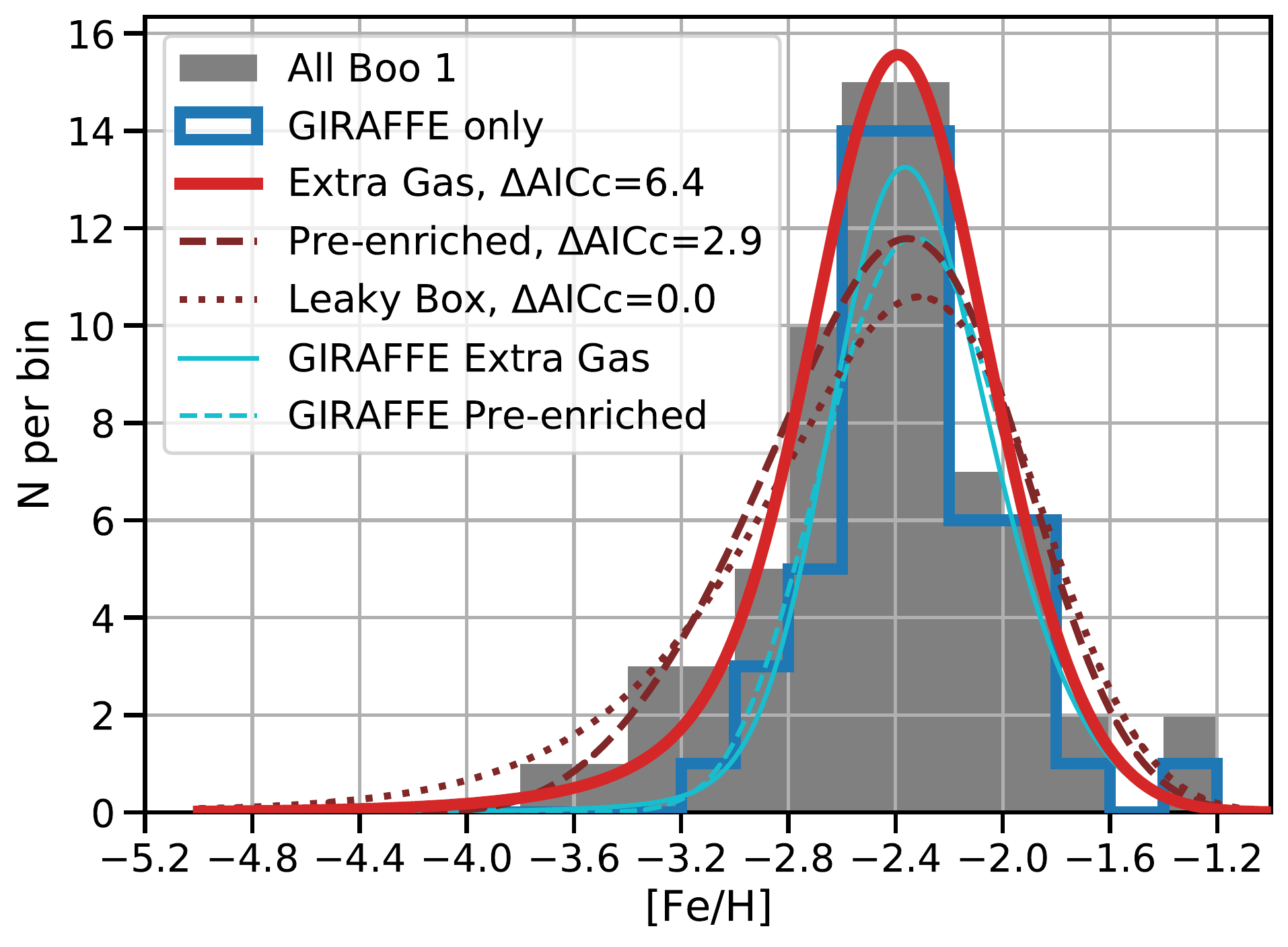}
    \caption{Bo{\"o}tes I MDF. The grey histogram depicts the MDF computed using VLT/GIRAFFE data in combination with the data analyzed by  \citet{nor2010} and \citet{lai2011}. The blue histogram depicts the MDF computed using only GIRAFFE data. The red and blue curves depict the model PDFs using combined data and VLT/GIRAFFE data, respectively. We find that the extra gas model provides the best fit to the combined data. }
    \label{fig:boo1mdf}
\end{figure}

We use dynamic nested sampling with \code{dynesty} \citep{dynesty} to determine the model parameters and posteriors.
The priors are log uniform for $p$ from $10^{-3}$ to $10^{-1}$ for all three models; uniform in $\feh_0$ from $-5$ to $-2$ for the pre-enriched model; and uniform in $M$ from 1 to 30 for the extra gas model.
The resulting posteriors are all well-behaved.
We find $\log p= -2.27 \pm 0.07$ for the leaky box model, $\log p = -2.33 \pm 0.07$ and $\feh_0 = -3.74 \pm 0.18$ for the pre-enriched box model, and $\log p = -2.32 \pm 0.05$ and $M = 4.5^{+3.2}_{-1.8}$ for the extra gas model.
To be consistent with previous similar analyses \citep{lai2011, kir2013, kir2020} we compare the models using the AICc.
Compared to the leaky box model, the pre-enriched model's AICc is 2.9 lower, and the extra gas model's AICc is 6.4 lower. \footnote{\revise{Note that in Figure \ref{fig:boo1mdf} we follow the convention used in \citet{kir2013} that a positive $\Delta$AICc corresponds to a lower AICc value compared to the reference AICc.}} The leaky box model is clearly disfavored, and the extra gas model is slightly preferred over the pre-enriched model.
Given that we have the full posterior, a Bayes factor could be more appropriate metric to compare these models, but doing so does not change these conclusions.

The new MDF and the results of our fit are shown in Figure~\ref{fig:boo1mdf}.
The model PDFs have been convolved with the median metallicity uncertainty of 0.22 dex. It is visually apparent that the extra gas model fits the histogram the best.
However, the strength of this conclusion relies on including all the literature data. Our VLT/GIRAFFE data alone rule out the pristine leaky box but cannot distinguish between the pre-enriched and extra gas models.
This is because the literature data are overall at slightly lower metallicities than the VLT data, increasing the size of the metal-poor tail compared to the peak of the MDF. While we have shifted the mean metallicity of the literature data to match our measurements, it is possible a residual metallicity offset remains, in which case the difference would not be significant. A homogeneous metallicity analysis of Bo{\"o}tes~I is needed to be certain.

Taking this result at face value, the MDF suggests that Bo{\"o}tes I was accreting gas, with $M=4.5^{+3.2}_{-1.8}$ times as much gas being accreted as stars being formed. A similar suggestion has been made by examination of detailed chemical abundances \citep{frebel16}, though in that case discrete merging events were responsible rather than continuous gas accretion as modeled here. 
Regardless, this interpretation is consistent with Bo{\"o}tes~I having formed in a similar way to the lowest mass classical dwarf galaxies \citep{kir2011,zhi2020}, emphasizing that relatively massive UFDs like Bo{\"o}tes I are an extension of ordinary galaxy formation to lower stellar masses \citep{tolstoy09,Simon2019ARA&A..57..375S}.
It remains to be seen if more populated MDFs in lower mass UFDs will continue this trend.
\subsection{Leo V Velocity Gradient} 
\label{sec:leov_gradient}
\citet{col2017} identified a velocity gradient with $\frac{dv}{dx}$ = $-$4.1$^{+2.8}_{-2.6}$ km s$^{-1}$ per arcmin and a preferred axis of $\phi$ = 123.6$^{+15.5}_{-29.6}$ deg. They argue that this is consistent with disruption caused by tidal interaction with the Milky Way.

We test whether there is a convincing velocity gradient present in the VLT data using a four-parameter model, as in \citet{col2017} and \citet{li2017}: 
\begin{equation}\label{eq:mcmc_gradient}
\medmuskip=-1.5mu
\thinmuskip=-1mu
\thickmuskip=-1mu
\log \mathcal{L} = -\frac{1}{2} \left [  \sum_{n = 1}^{N}\log (\sigma_{v_{\rm hel}}^2 + \sigma_{v_i}^2) +  \sum_{n = 1}^{N} \frac{(v_i - \rm\overline{v}_{hel} - \frac{dv}{d\chi}\chi_i)^2}{\sigma_{v_i}^2 + \sigma_{v_{\rm hel}}^2} \right ]
\end{equation}
The parameters of interest are $\rm\overline{v}_{hel}$, $\sigma_{v_\mathrm{hel}}$, velocity gradient $\frac{dv}{dx}$ and position angle of the gradient $\phi$. $\chi_i$ is the angular distance between the Leo V center ($\alpha_0, \delta_0$) and $i$-th star ($\alpha_i, \delta_i$) projected to the gradient axis at a position angle $\phi$:
\begin{equation}\label{eq:mcmc_chi}
\medmuskip=-1mu
\chi_i = (\alpha_i - \alpha_0)\cos(\delta_0)\sin(\phi) + (\delta_i - \delta_0)\cos(\phi).
\end{equation}

\begin{figure*}
\centering
\includegraphics[width=\textwidth]{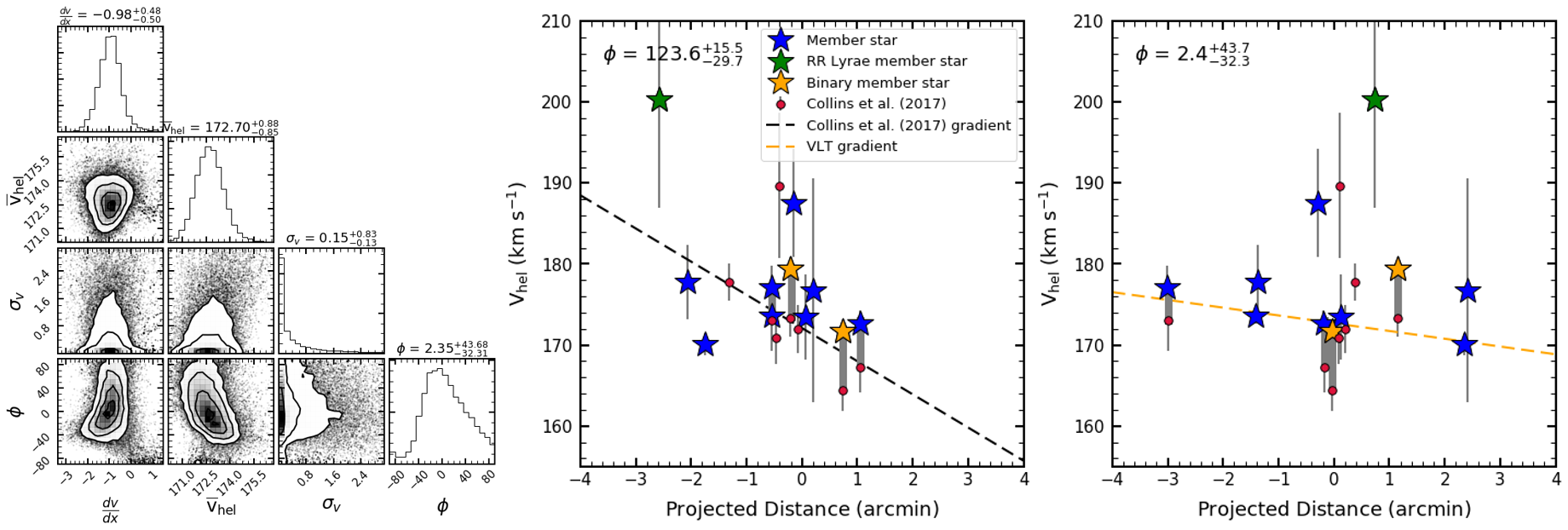}
\caption{Leo V velocity gradient. Left: Two-dimensional posterior probability distribution from an MCMC sampler using a four-parameter likelihood with the following parameters: systemic velocity $\overline{v}_\mathrm{hel}$ (km s$^{-1}$), velocity dispersion $\sigma$ (km $s^{-1}$), velocity gradient $\frac{dv}{dx}$ (km s$^{-1}$ per arcmin), and position angle $\phi$ (deg) over which the velocity gradient is maximised. Middle: The velocities as a function of projected distance along the gradient axis found by \citet{col2017}. Stars found in both the VLT and \citet{col2017} data are located at the same projected distance and are identified by a thick gray line. Our VLT velocity measurements for member stars are on average 4.8 $\pm$ 2.6 km s$^{-1}$ higher than the measurements from \citet{col2017}. The dashed line indicates the velocity gradient found by \citet{col2017}. The preferred kinematic axis is given in the top left. Right: The velocities as a function of projected distance along the gradient axis found in \S\ref{sec:leov_gradient}. We use the same symbols as used in the middle panel. The dashed line indicates the velocity gradient calculated from the VLT data. The preferred kinematic axis is given in the top left, and differs from the \citet{col2017} preferred axis by over 120 deg. 
  \label{vgradient}}
\end{figure*}

The posterior probability distributions derived from eight Leo V members (excluding binaries and RR Lyrae members) are displayed on the left side of Figure~\ref{vgradient}. We find $\frac{dv}{dx}$ = $-$0.98$^{+0.48}_{-0.50}$ km s$^{-1}$ per arcmin, which is consistent with zero within 2$\sigma$ uncertainty and is $\sim$4$\times$ smaller than the gradient calculated by \citet{col2017}. Additionally, we find $\phi$ = 2.4$^{+43.7}_{-32.3}$ deg, which is inconsistent with the preferred axis calculated by \citet{col2017} ($\phi$ = 123.6$^{+15.5}_{-29.7}$ deg). The weak velocity gradient from the VLT data at a very different preferred axis does not support the significant velocity gradient as seen in \citet{col2017}; instead, it shows that the inferred gradient from our work is likely due to the small sample used.
The discrepancy may also arise in part because \citet{col2017} observed a star (StarID-25 in their terminology) at a projected distance of $-$1.31 arcmin from the center of Leo V and with a radial velocity of 177.8 $\pm$ 2.3 km s$^{-1}$. This star contributed to their observed velocity gradient and was not observed with the VLT.
We demonstrate that, along the axis where \citet{col2017} find a gradient, the velocities of the VLT members from this work are relatively stable as a function of projected distance in the middle panel of Figure~\ref{vgradient}. Similarly, along the preferred axis from the VLT data (right panel of Figure~\ref{vgradient}), we do not observe an obvious trends in the \citet{col2017} data. 

The weak velocity gradient from the VLT data does not strongly indicate that Leo V is likely on the brink of dissolution, as suggested by \citet{col2017}. We also note that Leo V's orbit has a large pericenter (168$^{+12}_{-104}$ kpc) \citep{fri2018}, which is consistent with there being no tidal disruption. {\it Gaia} EDR3 will provide updated Leo V orbital information.
However, we identify additional nine stars greater than three half-light radii from the center of Leo V that we classify as VCNMs since they are far from the center of the galaxy. Though they exhibit other characteristics inconsistent with membership (e.g. high/low radial velocity, high metallicity, far from isochrone, etc.), if one were to be a member it may be evidence of tidal disruption. A dedicated observation including all known Leo V members  from previous literature (16 in total, including three RR Lyrae stars reported by \citet{med2017}) and this work is required to further examine the velocity gradient and possible tidal disruption feature in Leo V.

Finally, we caution that such a gradient study is performed on eight member stars which are selected  within 30 km s$^{-1}$ of the systematic velocity; this could in principle bias the inferred gradient if the gradient is larger than 10 km s$^{-1}$ per arcmin. However, there is no obvious evidence that such a large velocity gradient exists within the current member catalogue.

\section{Conclusions}
\label{sec:conclusion}
Using archived VLT spectroscopic data, \revise{we  present updated membership catalogues for three UFDs. } We summarize our key results for each UFD below: 

\begin{description}
  \item[$\bullet$ Bo{\"o}tes I] We identify 69 member stars in Bo{\"o}tes I. Using this membership catalogue, we calculate a systematic velocity of \revise{102.6$^{+0.7}_{-0.8}$ km s$^{-1}$, velocity dispersion of 5.1$^{+0.7}_{-0.8}$ km s$^{-1}$, systematic metallicity  of $-$2.34$^{+0.05}_{-0.05}$ dex and metallicity dispersion of 0.28$^{+0.04}_{-0.03}$ dex}. \revise{Potential binary stars were excluded from all kinematic calculations for Bo{\"o}tes I, Leo IV and Leo V.} We measure one member star with a metallicity of $-$1.22 $\pm$ 0.08 dex. When this star is excluded from metallicity calculations, we calculate a metallicity dispersion of 0.18$^{+0.04}_{-0.03}$ dex. \revise{We find weak evidence in support of the two-component kinematic model used by \citet{kop2011}, consistent with previous findings.} Combining the VLT/GIRAFFE data with data from \citet{nor2010} and \citet{lai2011}, we re-analyze the Bo{\"o}tes I MDF. We use three distributions from \citet{lai2011} (leaky box, pre-enriched leaky box and extra gas model) and find that the extra gas model (i.e. a model including infall of pristine gas while the galaxy was forming stars) best describes the MDF, suggesting that Bo{\"o}tes I formed in a similar way to low mass classical dwarf galaxies. \revise{In addition, we find strong evidence that one star (Boo1\_26) is a binary and find weak evidence indicating that three stars (Boo1\_61, Boo1\_111 and Boo1\_114) are binary. \citet{kop2011} had previously identified three of these stars (Boo1\_26, Boo1\_61 and Boo1\_111) as possible binaries.}
  \item[$\bullet$ Leo IV] We identify 20 member stars in Leo IV, including nine new members. Using this membership catalogue, we calculate a systematic velocity of \revise{131.6$^{+1.0}_{-1.2}$ km s$^{-1}$, velocity dispersion of 3.4$^{+1.3}_{-0.9}$ km s$^{-1}$, systematic metallicity  of $-$2.48$^{+0.16}_{-0.13}$ dex and metallicity dispersion of 0.42$^{+0.12}_{-0.10}$ dex}. This is the first time the velocity dispersion of Leo IV has been resolved at the 95\% confidence level. We measure one member star with a metallicity of $-$1.30 $\pm$ 0.15 dex. When this star is excluded from metallicity calculations, we are unable to resolve the Leo IV metallicity dispersion. In addition, we identify one new possible binary star \revise{(Leo4\_1039)} in Leo IV. 
  \item[$\bullet$ Leo V] We identify eleven member stars in Leo V, including four new members. Using this membership catalogue, we calculate a systematic velocity of \revise{173.1$^{+1.0}_{-0.8}$ km s$^{-1}$, systematic metallicity  of $-$2.29$^{+0.14}_{-0.17}$ dex and metallicity dispersion of 0.30$^{+0.14}_{-0.09}$ dex}. We also provide further evidence that one Leo V member star \revise{(Leo5\_1038)} is a binary, as suggested by \citet{mut2020}, and identify a new possible binary star \revise{(Leo5\_1034)}. We do not resolve the velocity dispersion when the two binaries are excluded. We identify a Leo V velocity gradient of $-$0.98$^{+0.48}_{-0.50}$ km s$^{-1}$ per arcmin, $\sim$4$\times$ smaller than the gradient calculated by \citet{col2017}. The gradient is consistent with zero within 2$\sigma$ uncertainty and is likely caused by the small sample size. \revise{This indicates that Leo V is not tidally disrupted.} Additionally, we calculate a preferred kinematic axis that differs from \citet{col2017} value by $\sim$120 deg.
\end{description}

Because Bo{\"o}tes I contains many ($>$50) known member stars, including or excluding a small number of plausible members or binary stars does not have a significant effect on the velocity or metallicity dispersion. However, for Leo IV and Leo V (in which only 10-20 members are identified), this could impact the final results. For example, the metallicity dispersion of Leo IV changes from 0.42$^{+0.12}_{-0.09}$ dex to unresolved when one plausible high metallicity member star is excluded, and the velocity dispersion of Leo V becomes unresolved when two binaries are removed from the calculation. For these faint UFDs, more comprehensive observations including all possible bright members are required to better constrain their kinematic and chemical properties.

We provide all spectroscopic measurements and membership results for Bo{\"o}tes I, Leo IV and Leo V in Table~\ref{tab:table_allstars}, with more details on the member stars of each galaxy in Tables~\ref{tab:table_Bootes_members}--\ref{tab:table_leov_members}. 

\revise{This is the first in a series of papers providing consistent, refined measurements of thirteen UFDs using public archival data from VLT/GIRAFFE.
Similar spectroscopic analyses for the remaining ten UFDs listed in Table~\ref{tab:table_ufd_list} will be presented in forthcoming papers. This series of work will largely improve the sample size of the known UFD member stars for refined kinematic and metallicity property studies on UFDs, provide multi-epoch observations to improve our understanding on the binary stars, and constrain the mass-metallicity relationship in these faintest galaxies.} 

\section{Acknowledgements}
S.J. and T.S.L. would like to thank Joshua Simon for helpful comments that significantly improved the quality of the paper. S.J. would also like to thank Joshua Frieman for providing the opportunity to work with T.S.L., as well as for comments on the paper.

S.J. is supported by the University of Chicago's Provost Scholar Award. 
T.S.L. is supported by NASA through Hubble Fellowship grant HST-HF2-51439.001, awarded by the Space Telescope Science Institute, which is operated by the Association of Universities for Research in Astronomy, Inc., for NASA, under contract NAS5-26555. 
A.B.P. is supported by NSF grant AST-1813881. B.M.P. is supported by an NSF Astronomy and Astrophysics Postdoctoral Fellowship under award AST-2001663.
A.P.J. is supported by a Carnegie Fellowship and the Thacher Research Award in Astronomy.

This work is based on observations collected at the European Organisation for Astronomical Research in the Southern Hemisphere under ESO programmes 82.B-0372(A), 185.B-0946(A) and 185.B-0946(B). 

This work has made use of data from the European Space Agency (ESA) mission Gaia (https://www.cosmos.esa.int/gaia), processed by the Gaia Data Processing and Analysis Consortium (DPAC, https://www.cosmos.esa.int/web/gaia/dpac/consortium). Funding for the DPAC has been provided by national institutions, in particular the institutions participating in the Gaia Multilateral Agreement.

The Legacy Surveys consist of three individual and complementary
projects: the Dark Energy Camera Legacy Survey (DECaLS; NOAO Proposal
ID \# 2014B-0404; PIs: David Schlegel and Arjun Dey), the
Beijing-Arizona Sky Survey (BASS; NOAO Proposal ID \# 2015A-0801; PIs:
Zhou Xu and Xiaohui Fan), and the Mayall z-band Legacy Survey (MzLS;
NOAO Proposal ID \# 2016A-0453; PI: Arjun Dey). DECaLS, BASS and MzLS
together include data obtained, respectively, at the Blanco telescope,
Cerro Tololo Inter-American Observatory, National Optical Astronomy
Observatory (NOAO); the Bok telescope, Steward Observatory, University
of Arizona; and the Mayall telescope, Kitt Peak National Observatory,
NOAO. The Legacy Surveys project is honored to be permitted to conduct
astronomical research on Iolkam Du'ag (Kitt Peak), a mountain with
particular significance to the Tohono O'odham Nation.

NOAO is operated by the Association of Universities for Research in
Astronomy (AURA) under a cooperative agreement with the National
Science Foundation.

The Legacy Survey team makes use of data products from the Near-Earth
Object Wide-field Infrared Survey Explorer (NEOWISE), which is a
project of the Jet Propulsion Laboratory/California Institute of
Technology. NEOWISE is funded by the National Aeronautics and Space
Administration.

The Legacy Surveys imaging of the DESI footprint is supported by the
Director, Office of Science, Office of High Energy Physics of the
U.S. Department of Energy under Contract No. DE-AC02-05CH1123, by the
National Energy Research Scientific Computing Center, a DOE Office of
Science User Facility under the same contract; and by the
U.S. National Science Foundation, Division of Astronomical Sciences
under Contract No. AST-0950945 to NOAO.

\bibliography{main}{}


\clearpage

\appendix
\section{Equivalent widths of M\lowercase{g}  I line at 8806.8 \AA} 
\label{sec:mg_line}
\begin{figure}
\centering
\plottwo{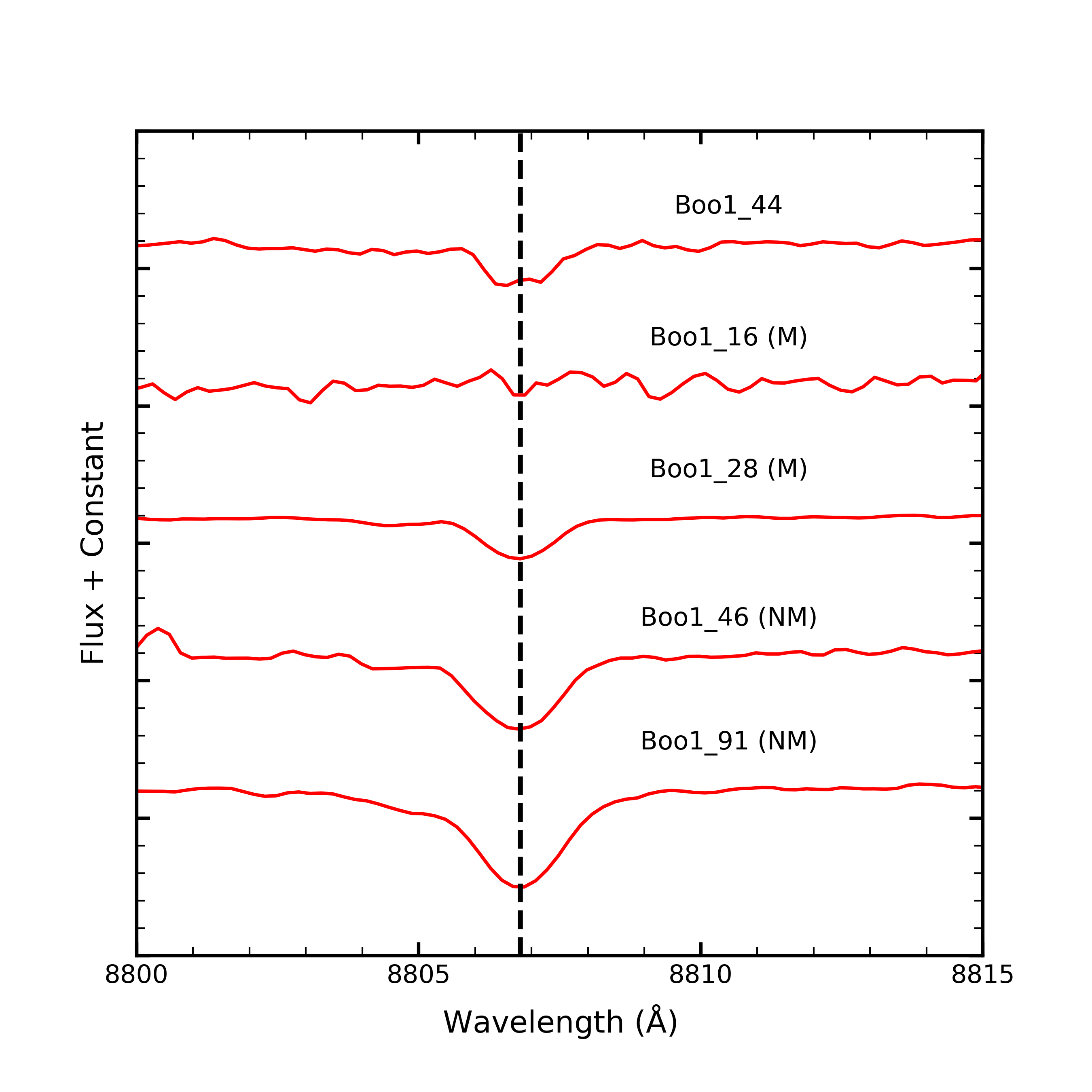}{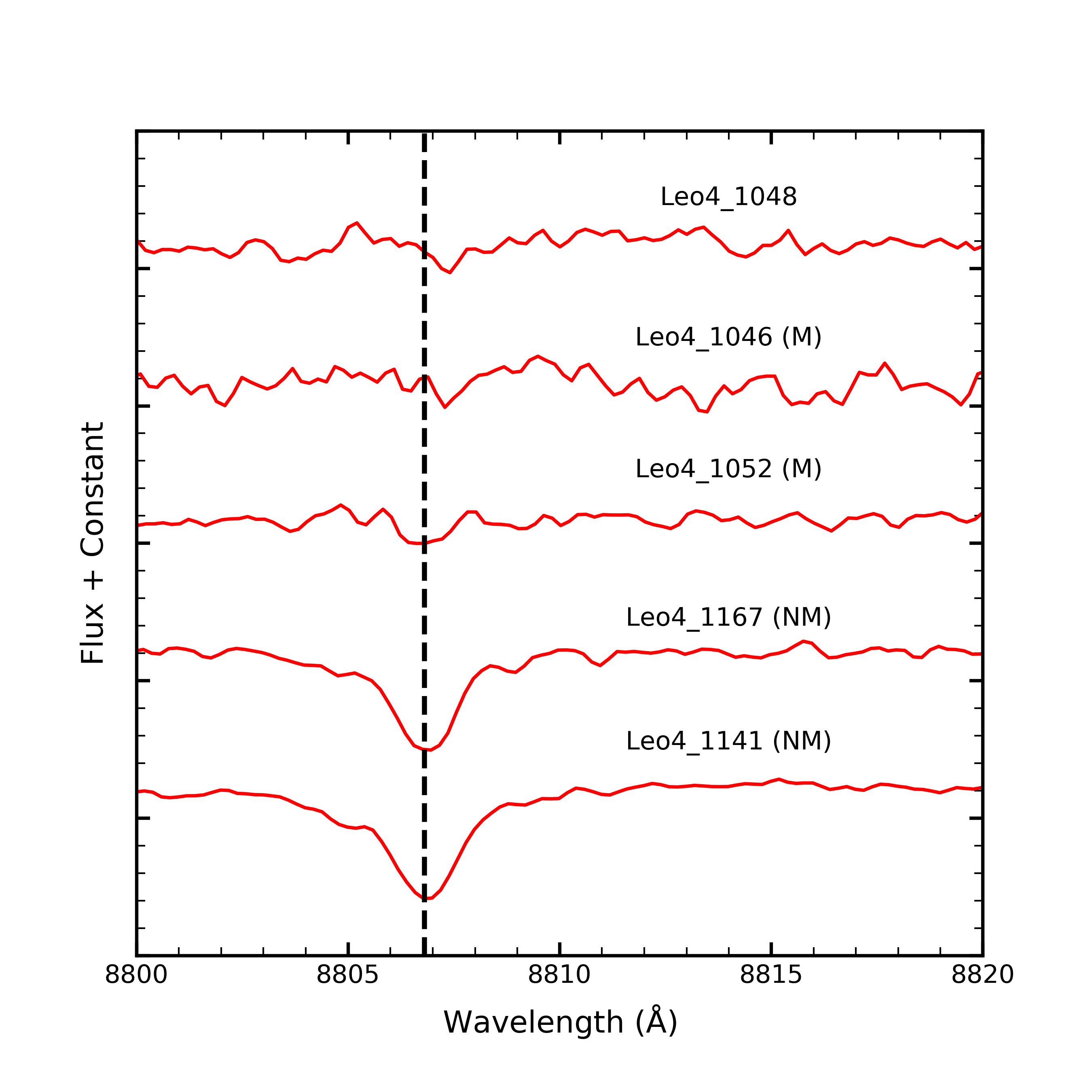}
\caption{Left: Restframe spectra of the high-metallicity star Boo1\_44 centered at the Mg I line, indicated by black dotted line. Spectra of Bo{\"o}tes I member (M) and non-member (NM) stars are shown for comparison. Right: Restframe spectra of the high-metallicity star Leo4\_1048. Spectra of Leo IV member (M) and non-member (NM) stars are shown for comparison.
  \label{fig:mgline}}
\end{figure}

As shown in \citet{bat2011} and \citet{bat2012}, the Mg I line at 8806.8 \AA~is gravity sensitive and can be used to discriminate foreground dwarf stars from the giant stars in the dwarf galaxies. We therefore compute the EWs of the Mg I line for all observed stars by integrating the flux over 6 \AA~around the central wavelength of the line, as was done in \citet{bat2011}, and report our measurements in Table~\ref{tab:table_allstars}. \revise{Due to the weakness of the Mg I line, it is only visible for bright stars with high S/N. For this reason, we do not use the line for general membership classification.} However, the Mg I line is useful for stars with uncertain membership, such as stars with high metallicity but consistent velocity and proper motion, e.g. Boo1\_44 (\feh = $-$1.22$\pm$0.08) in Bo{\"o}tes I and Leo4\_1048 (\feh = $-$1.30$\pm$0.15) in Leo IV. In Figure \ref{fig:mgline}, we show the spectra of these two stars centered at the Mg I line, along with several member and non-member stars for comparison. Although Boo1\_44 presents a weak Mg I line with an EW of 0.26 $\pm$ 0.02, the strength still places it in the ``giant" star region as defined by \citet{bat2012}. Leo4\_1048 does not show any obvious Mg I line. We therefore conclude that both stars are likely to be giant member stars rather than foreground dwarf stars.

\section{All Observed Stars}
\label{sec:allstars}
We present radial velocity and metallicity measurements (see \S\ref{sec:measurements}) for all observed stars in Table~\ref{tab:table_allstars}, in addition to membership probabilities and subjective membership classification (see \S\ref{sec:members}). A portion of the table is shown here to demonstrate its content. 
\begin{table}[t] 
\centering 

\notsotiny 
\caption{Measurements of all observed stars in Bo{\"o}tes I, Leo IV and Leo V. Only the first four lines are shown. The complete table is available online in a machine readable format. Column (1) is the star ID. Columns (2) and (3) correspond the coordinates and column (4) is the signal-to-noise ratio. Columns (5) and (6) present the radial velocity and CaT equivalent widths described in \S\ref{sec:measurements}, while column (7) provides the Mg equivalent widths. Column (8) corresponds to the probability calculated using a random forest classifier of the spectrum being good-quality (see \S\ref{sec:rv}). Column (9) presents the probability of the star being a member (see \S\ref{sec:members}) and column (10) provides the results of our subjective membership classification, with `-1' corresponding to non-member stars and `1' corresponding to member stars. } \label{tab:table_allstars} 
\hbox to\linewidth{\hfil\begin{tabular}{lllrrlrllr} 
\tablewidth{0pt} 
\hline 
\hline 
ID & RA (deg) & DEC (deg) & S/N & $v_\mathrm{hel}$ (km s$^{-1}$) & CaT EW & Mg EW  & GoodstarProb & MembershipProb & Member \\
\hline 
Boo1\_0 & 209.8390417 & 14.6017222 & 297.40 & -13.1$\pm$0.7 & 7.00$\pm$0.03 & 1.11$\pm$0.01 & 0.999 & 0.0 & -1 \\
Boo1\_1 & 209.844125 & 14.5501944 & 111.37 & 140.0$\pm$0.7 & 4.77$\pm$0.09 & 0.2$\pm$0.02 & 1.0 & 0.0 & -1 \\
Boo1\_2 & 209.8895833 & 14.4726667 & 7.26 & 118.0$\pm$5.2 & 0.99$\pm$0.20 & 0.05$\pm$0.36 & 0.999 & 0.901$^{+0.032}_{-0.043}$ & 1 \\
Boo1\_3 & 209.8932083 & 14.5047500 & 4.42 & 77.6$\pm$6.0 & 2.86$\pm$0.95 & 0.26$\pm$0.6 & 0.977 & 0.143$^{+0.103}_{-0.063}$ & 1 \\
\hline
\end{tabular}\hfil} 
\end{table}

\section{Mutual Stars}
\label{sec:mutualstars}
We identify several previously identified member stars in Leo IV and Leo V. Leo IV was previously observed by \citet{sim2007}. We compare the radial velocities of eleven common member stars in Figure~\ref{rv_comparison} and Table~\ref{tab:table_leoiv_previousmembers}. Leo V was previously observed by \citet{wal2009}, \citet{col2017} and \citet{mut2020}. We compare the radial velocities of seven common member stars in Figure~\ref{rv_comparison} and Table~\ref{tab:table_leov_previousmembers}.
\begin{table*}[ht]
\centering
\notsotiny
\centering
\caption{Properties of previously identified Leo IV member stars. The \citet{sim2007} IDs and radial velocity measurements are distinguished by $SG$. \label{tab:table_leoiv_previousmembers}}
\begin{tabular}{lcccccl}
\hline
\hline
 RA (deg) & Dec (deg) & ID & $v_\mathrm{hel}$ (km s$^{-1}$) & ID$_{SG}$ & v$_{SG}$ (km s$^{-1}$) & Comments\\
\hline
173.208875  &  $-$0.4446389  &  Leo4\_1087  &  128.0  $\pm$  2.5  &  383\_212  &  128.65  $\pm$  3.99 & \\
173.210375  &  $-$0.4978333  &  Leo4\_1057  &  127.8  $\pm$  4.8  &  383\_688  &  128.52  $\pm$  11.38 & \\
173.2110833  &  $-$0.5189722  &  Leo4\_1045  &  139.4  $\pm$  2.2 &  383\_262  &  137.8  $\pm$  5.32 & \\
173.2158333  &  $-$0.6271944  &  Leo4\_1080  &  139.9  $\pm$  3.5  &  383\_708  &  139.05  $\pm$  5.68 & \\
173.21775  &  $-$0.5382222  &  Leo4\_1039  &  134.2  $\pm$  2.9  &  383\_715  &  131.95  $\pm$  3.45 & Binary star\\
173.2232917  &  $-$0.5489722  &  Leo4\_1036  &  131.7  $\pm$  6.8  &  383\_738  &  124.91  $\pm$  5.76 & \\
173.2269167  &  $-$0.5530833  &  Leo4\_1037  &  136.2  $\pm$  2.8  &  383\_266  &  140.24  $\pm$  2.8 & \\
173.2325833  &  $-$0.55825  &  Leo4\_1041  &  132.4  $\pm$  9.2  &  383\_269  &  118.34  $\pm$  7.36 & RR Lyrae star\\
173.2372917  &  $-$0.5722222  &  Leo4\_1046  &  129.1  $\pm$  2.1  &  383\_393  &  135.12  $\pm$  2.89 & \\
173.2375  &  $-$0.5838611  &  Leo4\_1056  &  125.3  $\pm$  7.4  &  383\_391  &  126.22  $\pm$  5.02 & \\
173.2445833  &  $-$0.5805556  &  Leo4\_1052  &  131.2  $\pm$  1.0  &  383\_229  &  133.88  $\pm$  2.41 & \\
173.2558333  &  $-$0.5341944  &  Leo4\_1040  &  130.8  $\pm$  3.2  &  384\_278  &  135.79  $\pm$  3.51 & \\
\hline
\end{tabular}
\end{table*}
\begin{table*}[ht]
\centering
\notsotiny
\centering
\caption{Properties of previously identified Leo V member stars. The \citet{wal2009} IDs and radial velocity measurements are distinguished by $W$, the \citet{col2017} values are distinguished by $C$ and the \citet{mut2020} values are distinguished by $MP$. \label{tab:table_leov_previousmembers}} 
\begin{tabular}{lcccccccccl}
\hline
\hline
 RA (deg) & Dec (deg) & ID & $v_\mathrm{hel}$ (km s$^{-1}$) & ID$_W$ & v$_W$ (km s$^{-1}$) & ID$_C$ & v$_C$ (km s$^{-1}$) & ID$_{MP}$ & v$_{MP}$ (km s$^{-1}$) & Comments\\
\hline
 172.794125 & 2.2359722 & Leo5\_1038 & 179.4 $\pm$ 1.0 & L5$-$002 & 174.8 $\pm$ 0.9 & StarID$-$37 & 173.26 $\pm$ 2.3 & LeoV$-$6 & 176.1 $\pm$ 1.3, 169.5 $\pm$ 1.7 & Binary star\\
 172.805 & 2.2143333 & Leo5\_1037 & 172.7 $\pm$ 1.6 & L5$-$001 & 173.4 $\pm$ 3.8 & StarID$-$43 & 167.21 $\pm$ 3.1 & ..  & ... & \\
  172.7385833 & 2.1625556 & Leo5\_1069 & 177.1 $\pm$ 2.7 & .. & .. & StarID$-$17 & 173.02 $\pm$ 3.7 & ..  & ... & \\
 172.8002083 & 2.2165556 & Leo5\_1034 & 171.8 $\pm$ 0.9 & .. & .. & StarID$-$41 & 164.44 $\pm$ 2.5 & ..  & ... & Binary star\\
  172.7569167 & 2.1903056 & Leo5\_1046 & 173.6 $\pm$ 0.9 & L5$-$007 & 168.8 $\pm$ 1.6 & .. & .. & ..  & ... &  \\
 172.7672917 & 2.449 & Leo5\_1158 & 176.5 $\pm$ 4.5 & L5$-$057 & 179.2 $\pm$ 3.7 & .. & .. & ..  & ... & \\
 172.8087917 & 2.4434444 & Leo5\_1153 & 169.7 $\pm$ 3.1 & L5$-$052 & 165.6 $\pm$ 2.4 & .. & .. & ..  & ... & \\
\hline
\end{tabular}
\end{table*}
\end{document}